\documentclass[twocolumn]{aastex631}

\usepackage{graphicx}	
\usepackage{amsmath}	
\usepackage{amssymb}	
\usepackage{color}
\usepackage{algorithm}
\usepackage{booktabs}
\usepackage[noend]{algpseudocode}
\usepackage{pifont}
\usepackage{soul}
\usepackage{float}
\usepackage{multibib}
\newcites{Appendix}{Appendix References}
\usepackage{natbib}
\usepackage[flushleft]{threeparttablex}
\usepackage{mathrsfs}
\usepackage{tikz}
\usepackage{esint}
\usepackage[draft]{todonotes}
\usepackage{lipsum}
\usepackage{longtable}
\usepackage{pifont}
\usepackage{savesym}
\savesymbol{tablenum}
\usepackage{siunitx}
\restoresymbol{SIX}{tablenum}
\usepackage{placeins}
\usepackage{bm}

\newcommand{\appropto}{\mathrel{\vcenter{
  \offinterlineskip\halign{\hfil$##$\cr
    \propto\cr\noalign{\kern2pt}\sim\cr\noalign{\kern-2pt}}}}}

\renewcommand{\d}[1]{\ensuremath{\operatorname{d}\!{#1}}}

\DeclareMathOperator\D{\mathcal{D}}
\DeclareMathOperator\V{\mathcal{V}}
\newcommand{\M}{\mathcal{M}}

\DeclareMathOperator\s{\text{s}}
\DeclareMathOperator\f{\text{f}}

\DeclareMathAlphabet\mathbfcal{OMS}{cmsy}{b}{n}

\renewcommand{\v}{\mathit{v}}
\renewcommand{\b}{\mathit{b}}

\renewcommand{\u}{\bm{u}}
\renewcommand{\k}{\bm{k}}
\newcommand{\x}{\bm{x}}
\newcommand{\K}{\bm{K}}
\renewcommand{\bell}{\bm{\ell}}
\renewcommand{\P}{\mathcal{P}_{u}}

\newcommand{\It}{\mathbb{I}}
\newcommand{\T}{\mathcal{T}}
\newcommand{\bnab}{\bm{\nabla}}

\begin{document}

\correspondingauthor{\\
$^{\dagger}$Isabelle Connor: \href{mailto:iconnor@ucsc.edu}{iconnor@ucsc.edu} \\
$^{\ddagger}$James R. Beattie: \href{mailto:james.beattie@princeton.edu}{james.beattie@princeton.edu}}

\title[]{Cascading from the winds to the disk: \\ the universality of supernovae-driven turbulence in different galactic interstellar media}

\author[0009-0008-3986-657X]{Isabelle  Connor$^{\dagger}$}
    \affiliation{Department of Astronomy and Astrophysics, University of California, Santa Cruz, 1156 High Street, Santa Cruz, CA 96054} 

\author[0000-0001-9199-7771]{James R. Beattie$^{\ddagger}$}
    \affiliation{Canadian Institute for Theoretical Astrophysics, University of Toronto, 60 St. George Street, Toronto, ON M5S 3H8, Canada} 
    \affiliation{Department of Astrophysical Sciences, Peyton Hall, Princeton University, Princeton, NJ 08544, USA}

\author[0000-0001-7364-4964]{Anne Noer Kolborg}
    \affiliation{Department of Astronomy and Astrophysics, University of California, Santa Cruz, 1156 High Street, Santa Cruz, CA 96054} 

\author[0000-0003-2558-3102]{Enrico Ramirez-Ruiz}
    \affiliation{Department of Astronomy and Astrophysics, University of California, Santa Cruz, 1156 High Street, Santa Cruz, CA 96054}
    
\begin{abstract}
    Star-forming galaxies are in a state of turbulence, with one of the principle components of the turbulence sourced by the constant injection of momentum from supernovae (SNe) explosions. Utilizing high-resolution stratified, gravito-hydrodynamical models of SNe-driven turbulence with interstellar medium (ISM) cooling and heating, we explore how SNe-driven turbulence changes across different galactic conditions, parameterized by the galactic mass and potential, SNe-driving rate, and seeding functions. We show that even though the underlying ISM changes between starburst and Milky Way analogue models, the velocity fluctuations in the turbulence of both models, but not the kinetic energy fluctuations, can be normalized into a universal, single cascade, $du^2(k)/dk \propto k^{-3/2}$, where $u$ is the velocity and $k$ is the wavemode, indicating that the structure of the turbulence is robust to significant changes in the ISM and SNe seeding. Moreover, the cascades connect smoothly from the winds into the galactic disk, pushing the outer-scale of the turbulence, $\ell_{\rm cor}$, to over $\ell_{\rm cor} \approx 6 \ell_0$, where $\ell_0$ is the gaseous scale-height. By providing an analytical model for the sound speed spectrum, $dc_s^2(k)/dk$, in the weak-cooling, adiabatic limit, we show that it is the compressible turbulent modes, $u_c$, that control the volume-filling phase structure of the galactic disks in our models, with $dc_s^2(k)/dk \propto k^{-2} \propto du_c^2(k)/dk$. This may indicate that galactic turbulence does not only have highly-universal features across different galaxies, but also directly sets the volume-filling hot and warm phase structure of the underlying galactic ISM through turbulent compressible modes.
\end{abstract}


\section{Introduction} \label{sec:intro}
    Turbulence plays a critical role in shaping galactic structure and evolution via its influence on a variety of fundamental processes. It regulates star formation by creating both the over-dense regions necessary for stars to form, and the turbulent pressure that prevents immediate collapse \citep{Padoan2002,Krumholz2005,McKee2007,Hennebelle2011,MacLow2004,Federrath2012,Federrath2015_inefficient_SFR,Burkhart2018,Burkhart2019}. Furthermore, it drives metal mixing through turbulent diffusion \citep{Krumholz2018_metallicity_SF,Macias2018,Sharda2021_galactic_metallicity_modelling,Kolborg2022_metal_mixing_1,Kolborg2023_metal_mixing_2,Krumholz2025_metal_mixing_2}, and modulates cosmic ray transport through particle-wave interactions, intermittent structure generation, or the creation of intense inhomogeneities throughout the medium \citep{Socrates2008,Clark2015_cold_structures,Hu2021_cosmic_ray_modes,Kempski2022_cr_scattering,Xu2022_cosmic_ray_streaming,Ruszkowski2023_CRs_in_galaxies_review,Kempski2023_b_field_reversals,Sampson2023_SCR_diffusion,Beattie2022_ion_alfven_fluctuations,Sampson2025_CR_transport_CRMHD,Ewart2025_CR_multiphase_medium}.

    There are many sources of turbulence in the interstellar medium (ISM) of a galaxy, including galactic shear, stellar feedback, thermal instabilities, magneto-rotational instability, plasma accretion onto the galactic disk, and supernovae \citep{MacLow2004,Kim2002_MRI_galactic_disk,Elmegreen2004,Kim2006_galactic_spiral_shocks,Hill2012_SNe_driven_turb,Rosen2014,Federrath2016_brick,Krumholz2016_source_of_turb,Gallegos-Garcia2020,Sharda2021_driving_mode,Gerrard2023_LMC}. Among these, supernovae (SNe) are considered one of the primary components in a galaxy that sustain the turbulence \citep{MacLow2004,Hill2012_SNe_driven_turb,Gent_2017,2020Galax...8...56C}. SNe generate turbulence potentially through a variety of means, releasing energy and momentum into the ISM, influencing and regulating the multiphase plasma environment \citep{McKee1977_ISM,Guo2024_SNe_multiphase_ISM}. They contribute to initiating galactic outflows \citep[e.g.,][]{Martizzi2016,Fielding_winds,Dong2018_wind_review,Hu2019_Sne_driven_winds}, and provide ample velocity fluctuations to excite a turbulence cascade throughout an entire galaxy \citep{MacLow2004,Hill2012_SNe_driven_turb,Gent_2017,2020Galax...8...56C,Beattie2025_so_long_k41}, as we show below.

    Let us consider the energy from a core-collapse SN explosion, $E_{\rm SNe} \approx 10^{51}\,\rm{erg}$, which happens at a rate $\gamma_{\rm SNe} \approx 0.03 - 0.01 \, \rm{yr}^{-1}$ in our Galaxy \citep{Diehl2006_SNe_rate_MW}, leading to an energy flux rate given by $\dot{E}_{\rm SNe} \approx E_{\rm SNe}\gamma_{\rm SNe} \approx 3 \times 10^{41}\,\rm{erg}\,\rm{s}^{-1}$. The associated turbulent energy density can then be written as $e_{\rm turb} \approx (1/2)\rho_0 \left\langle u^2 \right\rangle$, where $\rho_0 \approx 1.67 \times 10^{-24}\,\rm{g}\,\rm{cm}^{-3}$ is the mean ISM density in our Galaxy, and $\left\langle u^2 \right\rangle^{1/2} \approx 10\,\rm{km\,s}^{-1}$ is the average integral root-mean-squared turbulent velocity \citep{Ferriere2020_reynolds_numbers_for_ism}. This provides a valuable estimate for the turbulent energy, $E_{\rm turb} \approx  e_{\rm turb}\mathcal{V}_{\rm ISM} \approx 6 \times 10^{53}\,\rm{erg}$, where $\V_{\rm ISM} \approx \pi R^2 \ell_0 \approx \pi (15\,\rm{kpc})^2 (0.1\,\rm{kpc}) \approx 7 \times 10^{33}\,\rm{cm}^{3}$. Here $\ell_0$ is the density scale-height and $R$ the radius of the Milky Way Galaxy. The turbulent eddy turnover time on $\ell_0$ is then $t_{\rm turb} \approx \ell_0 / \left\langle u^2 \right\rangle^{1/2} \approx 10^7\,\rm{yr}$, making the average energy flux rate from the turbulence $\dot{E}_{\rm turb} \approx E_{\rm turb}/t_{\rm turb} \approx 2 \times 10^{39}\,\rm{erg\,s}^{-1}$. Thus, with $\dot{E}_{\rm SNe} \gg \dot{E}_{\rm turb}$, there is sufficient energy flux from supernovae alone to sustain a galactic turbulence cascade.

    Foundational work from \citet{McKee1977_ISM} theorized that the entire phase structure of the Galactic ISM could be controlled and regulated by SNe and SNe-driven turbulence. Some of the first shearing box ISM simulations with SNe injection confirmed this to be a viable framework \citep[e.g.,][]{Korpi1999_SNe_ISM,Joung2006_stratified_box}, able to capture the correct volume-filling factors of the ISM phases \citep{deAvillez2005_ISM_phases_SNe}. Now, extremely high-resolution simulations of single SN show the entire bi-stable phase structure of the ISM in a single expanding remanent down to $0.01\,\rm{pc}$ resolutions \citep{Guo2024_SNe_multiphase_ISM}. On $\rm{kpc}$ scales, \citet{Hennebelle2014_SNe_driven_clustering} and \citet{Gatto2015_SNe_clustering} showed that the clustering of the SNe can change the nature of the volume of each of the ISM phases and the pressure equilibria of the mediums. However, to understand the nature of the underlying turbulence, and whether or not SNe can excite turbulence at all\footnote{Blastwaves can pass straight through one another without causing any significant turbulence at all \citep{Mee2006_blastwave_turbulence}.}, one has to study the Fourier power spectrum of velocities and energies.
    
    \citet{Padoan2016_supernova_driving}, \citet{Gent2022_multiphase_dynamo} and \citet{Beattie2025_so_long_k41} showed that there is a power law velocity and kinetic energy spectrum that resembles turbulence, albeit with deviations from the standard \citet{Kolmogorov1941} $du^2(k)/dk \propto k^{-5/3}$ model, following a shallower power law $du^2(k)/dk \propto k^{-3/2}$ \citep{Beattie2025_so_long_k41}, even in hydrodynamical simulations, suggesting that SNe-driven turbulence has a weakened nonlinearity compared to isotropic, incompressible turbulence. \citet{Balsara2004_SNe_turbulence_and_dynamo}, \citet{Padoan2016_supernova_driving}, \citet{Luibun2016_SNe_driving_modes} and  \citet{Beattie2025_so_long_k41} further showed that even though the energy and momentum is injected via SN blastwaves, the underlying plasma is dominated by incompressible velocity modes. \citet{Balsara2004_SNe_turbulence_and_dynamo} hypothesized that it was due to large-scale interaction between corrugated blastwaves having non-zero baroclinicity (similar to how cosmic shocks may generate magnetic fields through a Biermann battery; \citealt{Kulsrud1997_cosmic_batteries}). However, \citet{Beattie2025_so_long_k41} showed conclusively that the source of the incompressible modes is from the cooling layers in SNe remnants, where the baroclinicity at the interface between hot and warm plasma phases can be many orders of magnitude larger than in the background plasma (and the other incompressible mode generation pathways). Further detailed, spectral studies of SNe-driven turbulence in different galactic environments and with different numerical prescriptions, beyond ISM phase volume-filling factors and 1-point statistics, are critical for better understanding the nature of SNe-driven turbulence in different galaxies.  

    The cascade is not only shallower than \citet{Kolmogorov1941}, but it has inverse and forward directions in energy flux transfer. Using Helmholtz-decomposed transfer functions, \citet{Beattie2025_so_long_k41} showed that the SNe can indeed drive a number of cascades through triadic interactions between compressible, $u_c$, and incompressible, $u_s$, modes, with a mix of both inverse and forward cascades depending upon which triad of modes are interacting. \citet{Balsara2004_SNe_turbulence_and_dynamo} and \citet[][and others]{Gent2022_multiphase_dynamo} clearly showed that SNe-driven turbulence can excite a turbulent dynamo (e.g., \citealt{Kriel2022_kinematic_dynamo_scales,Kriel2023_fundamental_scales_II,Beattie2025_compressible_dynamo}), implying that not only is the phase structure of the interstellar plasma maintained by turbulent motions, but so may be the galactic small-scale and large-scale dynamo \citep{Beck_2013_Bfield_in_gal}, the process that maintains the amplitude and structure of the magnetic field in our Galaxy and others \citep{Clark2019,Surgent2023_magnetic_fields_in_galaxies}. This makes SNe-driven turbulence a fundamental and important aspect to study in its own right.

    In this study, we investigate how the clustering of SNe and the characteristics of the underlying galaxy model --  parameterized by the SNe explosion rate, the mass, and the depth of the gravitational potential -- affect the nature of the turbulence. We conduct this investigation by examining the velocity, kinetic energy, and sound speed spectra. While previous research has explored these aspects at the level of phase diagrams and distribution functions of thermodynamic properties \citep[e.g.,][]{Hennebelle2014_SNe_driven_clustering, Gatto2015_SNe_clustering},  the scale-dependent spectra analysis we conduct has not before been studied. We aim to address whether the underlying turbulence varies in different galaxies, or changes in response to different SNe seeding prescriptions.

    We demonstrate that even with significant alterations in the phase structure of the ISM, the gravitational potential of the disk and halo, the mass, the size of the disk, the SNe driving rate, and the inherent clustering of SNe, the resulting spectra either remain invariant (velocity), or smoothly vary into one another (kinetic). Specifically, the turbulence forms a consistent velocity power spectrum, \( u^2(k) \propto k^{-3/2} \), as the energy moves from the hot winds into the warm phases of the ISM within the disk. Moreover, the kinetic energy derived from simulations dominated by the disk (warm-phase) and those dominated by winds (hot-phase) can be seamlessly transformed into a single spectrum that connects these different phases across the simulations. Overall, these findings suggest that despite significant changes in the underlying phase structure of the galaxy, the fundamental turbulence in the velocity remains universal.

    This study is organized as follows. In \autoref{sec:numerics} we describe the key aspects of the multiphase gravito-hydrodynamical fluid models we use to simulate a section of a galactic disk, including the key parameters controlling the differences between each of the models (SNe-driving rate, gravitational potential and SNe seeding prescription). In \autoref{sec:ISM_diversity} we show the key ISM statistics coming from each of the simulations, including the vertical profiles of density and turbulence, the ISM phase diagrams, and the mass density distribution functions. Next, in \autoref{sec:power_spectrum} we show and discuss the Fourier power spectrum for a variety of quantities, including the velocity, compressible and incompressible velocity modes, the kinetic energy and the sound speed. Finally, in \autoref{sec:conclusions} we conclude, summarize and itemize the key results in this study.
    
\begin{deluxetable*}{lcccccccc}\label{tab:ics}
\tablecaption{Initial conditions for each of the four unique models.}
\tablehead{
\colhead{Model} & 
\colhead{$L/\rm{kpc}$} & 
\colhead{$2\pi G\Sigma_*/\rm{kpc}\;\rm{Myr}^{-2}$} & 
\colhead{$(4/3)\pi G\rho_{\rm h}\;\rm{Myr}^{-2}$} & 
\colhead{$z_0/\rm{pc}$} & 
\colhead{$\rho_{\rm init}/\rm{g\;cm}^{-3}$} &
\colhead{$T_{\rm init}/\rm{K}$} &
\colhead{$\gamma_{\rm SNe}/10^{-4}\rm{yrs}$} &
\colhead{$z_{\rm{SNe}}/\rm{pc}$}\\
\colhead{(1)} & \colhead{(2)} & \colhead{(3)} & \colhead{(4)} & \colhead{(5)} & \colhead{(6)} & \colhead{(7)} & \colhead{(8)} &\colhead{(9)}
}
\startdata
MWSC & $1.0$  & $1.42 \times 10^{-3}$ & $5.49 \times 10^{-4}$ & 180 & $2.08 \times 10^{-24}$ & $7670$ & $0.1$ & -\\
MWFX & $1.0$  & $1.42 \times 10^{-3}$ & $5.49 \times 10^{-4}$ & 180 & $2.08 \times 10^{-24}$ & $7670$ & $0.1$ & $160$\\
SBSC & $1.0$  & $1.26 \times 10^{-2}$ & $4.87 \times 10^{-3}$ & 180 & $3.47 \times 10^{-23}$ & $7670$ & $3.0$ & - \\
SBFX & $1.0$  & $1.26 \times 10^{-2}$ & $4.87 \times 10^{-3}$ & 180 & $3.47 \times 10^{-23}$ & $7670$ & $3.0$ & $80$\\
\hline\\[-1.25em]
\enddata
\tablecomments{\textbf{Column (1)}: Numerical model label, where MW means Milky Way analogue, SB means starburst analogue, SC refers to the self-consistent SNe seeding prescription, \autoref{sssec:sc}, and FX refers to the fixed seeding prescription, \autoref{sssec:sc}. \textbf{Column (2)}: Length of the cubic
simulation domain. \textbf{Column (3)}: Stellar disk component of $\phi(z)$ \autoref{eq:pressure_equilibrium}. \textbf{Column (4)}: Dark matter halo component of $\phi(z)$. \textbf{Column (5)}: Stellar disk scale height. \textbf{Column (6)}: Initial density. \textbf{Column (7)}: Initial temperature. \textbf{Column (8)}: Global, volumetric supernova driving rate. \textbf{Column (9)}: $z$ threshold for supernova-driving boundary (FX models only; \autoref{eq:distfx}). }
\end{deluxetable*}

\section{Numerical Methods \& Models}\label{sec:numerics}
    In this section, we summarize the fluid model and simulation parameters used to investigate the nature of interstellar turbulence driven by different supernova seeding schemes within varying galactic models. Our study focuses on two galaxy models, which are detailed in \autoref{ssec:code&conditions}. One model is designed to resemble the present-day Milky Way (MW), while the other represents a starburst galaxy (SB), chosen to approximate the MW and ULTRA-MW models discussed in \citet{Martizzi2016}.

\subsection{Supernovae-driven, gravito-hydrodynamical fluid model}
    We model sections of a galactic disk using the basic set-up described in \citet{Martizzi2016}, \citet{Kolborg2022_metal_mixing_1} and \citet{Beattie2025_so_long_k41}. For our ISM simulations, we solve the three-dimensional, compressible Euler equations in a static gravitational field with mass, momentum and energy sources from stochastic supernova events \citep[point sources of mass, energy and momentum based on detailed 1D models;][]{Martizzi2015}. To solve the fluid model we use the \textsc{ramses} code \citep{Teyssier2002_ramses}, employing the monotonic upstream-centered scheme for conservation laws (MUSCL) scheme and a HLLC Riemann solver. The model is
    \begin{align}
    \dfrac{\partial \rho}{\partial t} + \bnab \cdot \left( \rho \u \right) = {}& \dot{n}_{\rm SNe} M_{\rm ej}, \label{eq:mass_conservation} \\
    \dfrac{\partial \rho \u}{\partial t} + \bnab \cdot \left( \rho \u \otimes \u + P \mathbb{I} \right) = & - \rho \bnab \phi
    +\dot{n}_{\rm SNe} \bm{p}_{\rm SNe}(Z, n_{\rm H}), & \label{eq:momentum_conservation} \\
    \dfrac{\partial \rho e}{\partial t} + 
    \bnab \cdot \left[\rho \left( e + P \right) \u \right] =& - n_{\rm H}^2 \Lambda - \\
    \rho \u \cdot \bnab \phi
     + \dot{n}_{\rm SNe}\bigg[E_{\rm th, SNe}(&Z, n_{\rm H}) + \frac{p^2_{\rm SNe}(Z, n_{\rm H})}{2 (M_{\rm ej} + M_{\rm swept})} \bigg], \label{eq:energy_conservation}  \\
    e = & \epsilon + \frac{u^2}{2}, \,\, P = (\gamma - 1) \rho \epsilon, \label{eq:EOS}
    \end{align}
    where $\otimes$ is the tensor product, such that $\u\otimes\u = u_i u_j$. $\u$ is the gas velocity and $\rho$ is the mass density. $\bnab\phi$ describes the static gravitational potential, the details of this potential are described in \autoref{ssec: phi}. $P$ is the scalar pressure, and $\mathbb{I}$ is the unit tensor, $\delta_{ij}$. $\rho e$ is the total energy composed of both the gas kinetic energy, $\rho u^2/2$, and the internal energy, $\epsilon$, which is related to $P$ via, $P = (\gamma - 1) \rho \epsilon$, where we use $\gamma = 5/3$ as the adiabatic index for a monoatomic gas. $\dot{n}_{\rm SNe}$ is the volumetric rate of SNe explosions, which changes for the two seeding schemes (details in \autoref{ssec:Seeding}), $\bm{p}_{\rm SNe}$, $M_{\rm ej}$ and $E_{\rm{th,SNe}}$ are the radial momentum, ejecta mass and thermal energy of each SNe. $M_{\rm swept}$ is the mass of ISM material swept up by the SNe shock wave. $Z$ is the metallicity of the ambient medium, which we set to $Z = Z_{\odot}$ for simplicity. Further details of the SNe driving are discussed in \autoref{ssec:Seeding}. $\Lambda$ is the net-cooling function, encompassing both heating and cooling terms, summarized in \autoref{ssec:cooling} (see \citealt{Beattie2025_so_long_k41} for details). $n_H$ is the number density of hydrogen. Our goal with this setup is to perform controlled studies of the coupling between SNe and turbulence in the ISM, not to implement the most complex and complete physics possible. For a detailed discussion of the limitations our numerical model we refer to \S7.2 in \citet{Beattie2025_so_long_k41}.

\subsubsection{Gravitational potential} \label{ssec: phi}
    Our simulations use a two-component (stellar disk and dark matter halo) static gravitational potential, $\phi(z)$. It is parameterised with a stellar disk of scale height $z_0$, infinite thin disc of stellar surface density $\Sigma_*$, and a spherical dark matter halo of density $\rho_{\rm halo}$,
    \begin{align}\label{eq:phi}
        \phi(z)=2\pi G\Sigma_*\Big(\sqrt{z^2+z_0^2}-z_0\Big)+\frac{2\pi G\rho_{\rm halo}}{3}z^2
    \end{align}
    with components from the disk, $2\pi G\Sigma_*z/\sqrt{z^2+z_0^2}$, and dark matter halo, $(4/3)\pi G\rho_{\rm halo}z$, respectively \citep{KuijkenGilmore}. Our simulations neglect the self-gravity of the gas, which would be important for a full star formation simulation. We change the disk and halo potentials to parameterize between our Milky Way and starburst-type galactic models. The values for the components for each of the models are tabulated in Table~\ref{tab:ics}. 

\subsubsection{Initial \& boundary conditions} \label{ssec:code&conditions}
    The simulations are initialized in hydrostatic equilibrium with the static gravitational potential, $\bnab \cdot (P \It) = - \rho \bnab \phi$, at $T_{\rm init} = 7,\!670\,\rm{K}$. For our MW model, the gas density in the disk mid-plane at this time is $\rho_{\rm init} = \SI{2.1e-24}{g cm^{-3}}$, yielding a constant pressure $P_{\rm init} = \SI{2.2e-12}{Ba}$ throughout the domain. The disk mid-plane density for MW is chosen such that the gas surface density, $\Sigma_{\rm gas} = \SI{5}{M_\odot / pc^2}$, is similar to that of the present-day solar neighborhood value (\citealp{McKee_solarneigh}; \citealp[see also][]{Martizzi2016}). Our SB model has the initial midplane gas-density (and depth of the gravity potential) and supernova rate increased by approximately an order magnitude compared to the MW model, resulting in strong feedback and winds. For this setup, we have $\rho_{\rm init} = \SI{3.5e-23}{g cm^{-3}}$, leading to $P_{\rm init} = \SI{3.7e-11}{Ba}$, and $\Sigma_{\rm gas} = \SI{50}{M_\odot / pc^2}$, analogous to a galaxy on the low-end of metal-poor starbursts (see \citealp{Kennicutt_MW_SB}). In all models the gas velocity is initialized $|\u| = 0$, with no initial velocity perturbations. 

    We choose a cubic simulation domain with length, $L = \SI{1000}{pc}$, allowing us to explore the large-scale ISM properties of the disk, and the disk-wind connection in the turbulence. The simulations have periodic boundary conditions on the four sides perpendicular to the disk mid-plane and outflow boundaries on the top and bottom boundaries. The outflow boundaries suppress any inflowing material from the ghost cells. The domain is discretized using a regular Cartesian grid of up to $N_{\rm grid} = 512$ cells for each $L$. For second-order spatial reconstruction methods such as the one used to solve our model, the numerical diffusive effects influence $\approx 10\d{x}$ \citep{Shivakumar2025_numerical_dissipation,Beattie2025_compressible_dynamo}, where $\d{x} = L/N_{\rm grid}$, which means that at $N_{\rm grid} = 512$ we properly resolve roughly $\SI{50}{pc} \lesssim \ell \lesssim \SI{1000}{pc}$ in our simulations, i.e., the large-scale properties of the ISM. 

\begin{deluxetable*}{cccccccc}\label{tb:results}
\tablecaption{Midplane, scale-height and Mach number statistics for each of the simulations, averaged over the stationary state.}
\tablehead{
\colhead{Model} &
\colhead{$\ell_{\rm0}$ /pc} &
\colhead{$t_{\rm0}$ /Myr} &
\colhead{$\mathcal{M}$} &
\colhead{$\rho_{\rm0}$ /g cm$^{-3}$} & 
\colhead{$P_{\rm0}$/erg cm$^{-3}$} & 
\colhead{$T_{\rm0}$/K} &
\colhead{$\sigma_{\rm z}^2$/kms$^{-1}$}\\
\colhead{(1)} & \colhead{(2)} & \colhead{(3)} & \colhead{(4)} & \colhead{(5)} & \colhead{(6)} & \colhead{(7)} & \colhead{(8)}
}

\startdata
MWSC128 & $124.6\pm0.9$ & $3.6\pm0.8$ & $1.9 \pm 0.2$ & $(3.7\pm0.3)\times10^{-24}$ & $(1.9\pm0.2)\times10^{-12}$ & $3700\pm215$ & $7.27\pm0.05$\\
256 & $125.9\pm0.6 $ & $3.8\pm0.8$ & $1.9 \pm 0.1$ & $(3.6\pm0.2)\times10^{-24}$ & $(1.8\pm0.3)\times10^{-12}$& $3500\pm476$ & $7.34\pm0.03$\\
512 & $122.2\pm0.5 $ & $4.1\pm0.9 $ & $1.8\pm0.1$ & $(3.6\pm0.3)\times10^{-24}$ & $(1.7\pm0.2)\times10^{-12}$ & $3300\pm346$ & $7.14\pm0.03$\\
MWFX128 & $124.1\pm0.8 $ & $4.1\pm1.4 $ & $1.8 \pm 0.1$ & $(3.7\pm0.3)\times10^{-24}$ & $(1.9\pm0.2)\times10^{-12}$ & $3700\pm208$ & $7.25\pm0.04$\\
256 & $121.1\pm0.6 $ & $3.6\pm0.9$ & $1.7 \pm 0.1$ & $(3.8\pm0.2)\times10^{-24}$ & $(1.8\pm0.2)\times10^{-12}$ & $3500\pm247$ & $7.09\pm0.03$\\
512 & $117.7\pm0.5$  & $3.9\pm0.8$  & $1.7\pm0.1$ & $(3.8\pm0.2)\times10^{-24}$ & $(1.8\pm0.2)\times10^{-12}$ & $3300\pm250$ & $6.90\pm0.03$\\
SBSC128 &$44.7\pm0.2$ & $0.74\pm0.07$ & $11\pm50$ & $(4.6\pm0.1)\times10^{-23}$ & $(2.4\pm0.1)\times10^{-11}$ & $3690\pm70$ & $8.13\pm0.03$\\
256 & $41.5\pm0.1$ & $0.72\pm0.03$ & $41\pm10^3$ & $(4.9\pm0.2)\times10^{-23}$ & $(2.3\pm0.1)\times10^{-11}$ & $3440\pm88$ & $7.54\pm0.02$ \\
512 & $42.7\pm0.1$ & $0.77\pm0.02$ & $1.9\pm0.1$ & $(4.8\pm0.1)\times10^{-23}$ & $(2.2\pm0.1)\times10^{-11}$ & $3300\pm127$ & $7.77\pm0.02$ \\
SBFX128 &$44.2\pm0.2$ & $0.31\pm0.01$ & $735\pm6000$ & $(5.7\pm0.3)\times10^{-23}$ & $(2.8\pm0.2)\times10^{-11}$ & $3590\pm70$  & $8.03\pm0.04$\\
256 & $42.9\pm0.2 $  & $0.36\pm0.03$ & $2.9\pm0.3 $ & $(4.8\pm0.2)\times10^{-23}$ & $(2.1\pm0.1)\times10^{-11}$ & $3170\pm48$ & $7.80\pm0.03$ \\
512 & $42.4\pm0.1$ & $0.37\pm0.02$ & $2.0\pm0.2$ & $(4.8\pm0.2)\times10^{-23}$ & $(2.1\pm0.1)\times10^{-11}$ & $3170\pm48$ & $7.71\pm0.02$  \\
\hline\\[-1.25em]
\enddata
\tablecomments{All reported values in this table are computed and averaged over the statistically stationary state of the SNe-driven turbulence (\autoref{ssec:stationarity}), unless otherwise stated. The $1\sigma$ stated in the table encapsulate the physical variation of the quantities in the stationary state. \textbf{Column (1)}: the model label, with an additional suffix for the linear grid resolution of the simulation (e.g., MWFX512 means Milky Way analogue, fixed SNe seeding prescription, discretized on a $512^3$ domain). \textbf{Column (2)}: the gaseous scale height fit via a pressure-balanced isothermal atmospheric model based on our $\phi$, \autoref{eq:isothermal_profile}. \textbf{Column (3)}: the turbulent turnover time, \autoref{eq:t0}, on $\ell_0$. \textbf{Column (4)}: The turbulent Mach number, \autoref{eq:mach}. \textbf{Column (5)}: the midplane mass density. \textbf{Column (6)}: the midplane pressure. \textbf{Column (7)}: the midplane temperature. \textbf{Column (8)}: the isothermal dispersion supporting the disk component of each of the galaxy models, fit using our isothermal profile model \autoref{eq:isothermal_profile}. It has contributions from both the turbulence, $P_{\rm turb} \approx 1/2 \bnab u^2$ and the underlying thermal fluctuations. }
\end{deluxetable*}

\subsubsection{Heating and cooling} \label{ssec:cooling}
    We implement the heating and cooling model comprehensively described in \citet{Beattie2025_so_long_k41}, which is the same for all simulations. To summarize, based on \citet{Theuns1998_Cooling} and \citet{SutherlandDopita}, we solve a time-dependent HI, HII, HeI, HeII, HeIII and free electron chemical network at each integration step, in each grid cell based on density and temperature, using microphysical prescriptions that include photoheating and a range of cooling processes: collisional ionization, recombination, dielectronic recombination, collisional excitation, Bremsstrahlung and metal line cooling at solar metallicity. At $10^4\,\rm{K}\lesssim$ $T$ $\lesssim 10^8\,\rm{K}$, cooling is dominated by metal line emission from SNe, collisional excitation and ionization, and at $T\gtrsim10^7\,\rm{K}$ bremsstrahlung dominates \citep{Karpov2020}. Since we focus primarily on the large-scales due to our resolution limits, to save the computational cost, most cooling terms truncate at $T\approx10^4\,\rm{K}$, so there is no stable cold $T\approx 100\,\rm{K}$ phase. The cooling below this limit (the simulations form dilute gas with $T\lesssim 10^2\,\rm{K}$) is due to macroscopic processes, like adiabatic expansion of the disk (see \autoref{fig:mask}). At this scale, attempts to model cold phase gas (which condense in $\lesssim (0.1-10)\,\rm{pc}$ clumps \citep{Ferriere2020_reynolds_numbers_for_ism}, would be unresolved and unlikely to have a significant effect on our results. As will be seen in \autoref{sec:ISM_diversity}, our model is well-suited to capture the large-scale warm (WNM \& WIM) and hot (HIM) phases of the ISM, i.e., the volume-filling properties of the disk and wind connection in the ISM.

\subsection{Supernova seeding prescriptions}\label{ssec:Seeding}
    The drivers of the turbulence in this study are core-collapse SNe explosions. Thermal energy, mass, and momentum are injected using the sub-grid model from \citet{Martizzi2015,Martizzi2016}, which accounts for unresolved SNe evolution. The injections happen over a region of size $\ell_{\rm{inj}}=2\,\rm{dx}$ which is fixed for all resolutions (see \autoref{app:convergence} and \citealt{Beattie2025_so_long_k41} for spectra convergence tests and justification). The total energy, $E_{\rm{SNe}}=10^{51}\,\rm{erg}$, is partitioned based on the ambient $\rho$ and $Z$ in the local medium, into radial momentum, $p_{\rm{SNe}}$, and thermal energy, $E_{\rm{th,SNe}}$. The SNe each eject $M_{\rm{ej}}=6.8\,M_\odot$ new material into the ISM and continuously sweep material $M_{\rm{swept}}$ through the medium as the SNR expands. We compare two different seeding mechanisms for each galaxy type to probe their effects on the turbulence, following the ``fixed" (FX) and ``self-consistent" (SC) seeding prescriptions introduced in \citet{Martizzi2016}, and detailed in the following section.

\subsubsection{Fixed seeding (FX)}\label{sssec:fx}
    In the FX scheme, SNe are uniformly seeded in space and time within a fixed height, $z_{\rm{SNe}}$. The SNe have an equal probability of exploding within this region, and zero probability outside of the region,
    \begin{align}\label{eq:distfx}
        p(z)=
        \begin{cases}
        1/(2z_{\rm{SNe}}), & |z|\leq z_{\rm{SNe}},\\
        0, & |z| > {z_{\rm{SNe}}},
        \end{cases}
    \end{align}
    where $z_{\rm{SNe}}=160\,\rm{pc}$ for MW, and $z_{\rm{SNe}}=80\,\rm{pc}$ for SB. This follows the method in \citet{Kolborg2023_metal_mixing_2} where $z_{\rm{SNe}}=2z_{\rm{eff}}$, $z_{\rm{eff}}$ being the initial gaseous scale height analytically derived from initial conditions (not including turbulence or SNe contributions).
    
    The global volumetric rate of SNe, $\dot{n}_{\rm{SNe}}$, is
    \begin{align}\label{ratefx}
        \dot{n}_{\rm{SNe,FX}}=\frac{\dot{\Sigma}_*}{z_{\rm{SNe}}100\rm{M}\odot}
    \end{align}
    where $\dot{\Sigma}_*\propto\Sigma_{\rm{gas}}^{1.4}$ is the surface density of star formation \citep{Kennicutt1998_SFR,Kennicutt2007_obsSFR}. This simple seeding prescription allows SNe to explode with uniform probability inside of our steady-state gaseous disk measured in \autoref{ssec:scaleheight}. Given this mechanism is simple and not linked in detail to a particular physical model, it is useful for comparison to the following more comprehensive self-consistent SNe seeding method.

    \begin{figure}[htbp]
        \centering
        \includegraphics[width=0.9\linewidth]{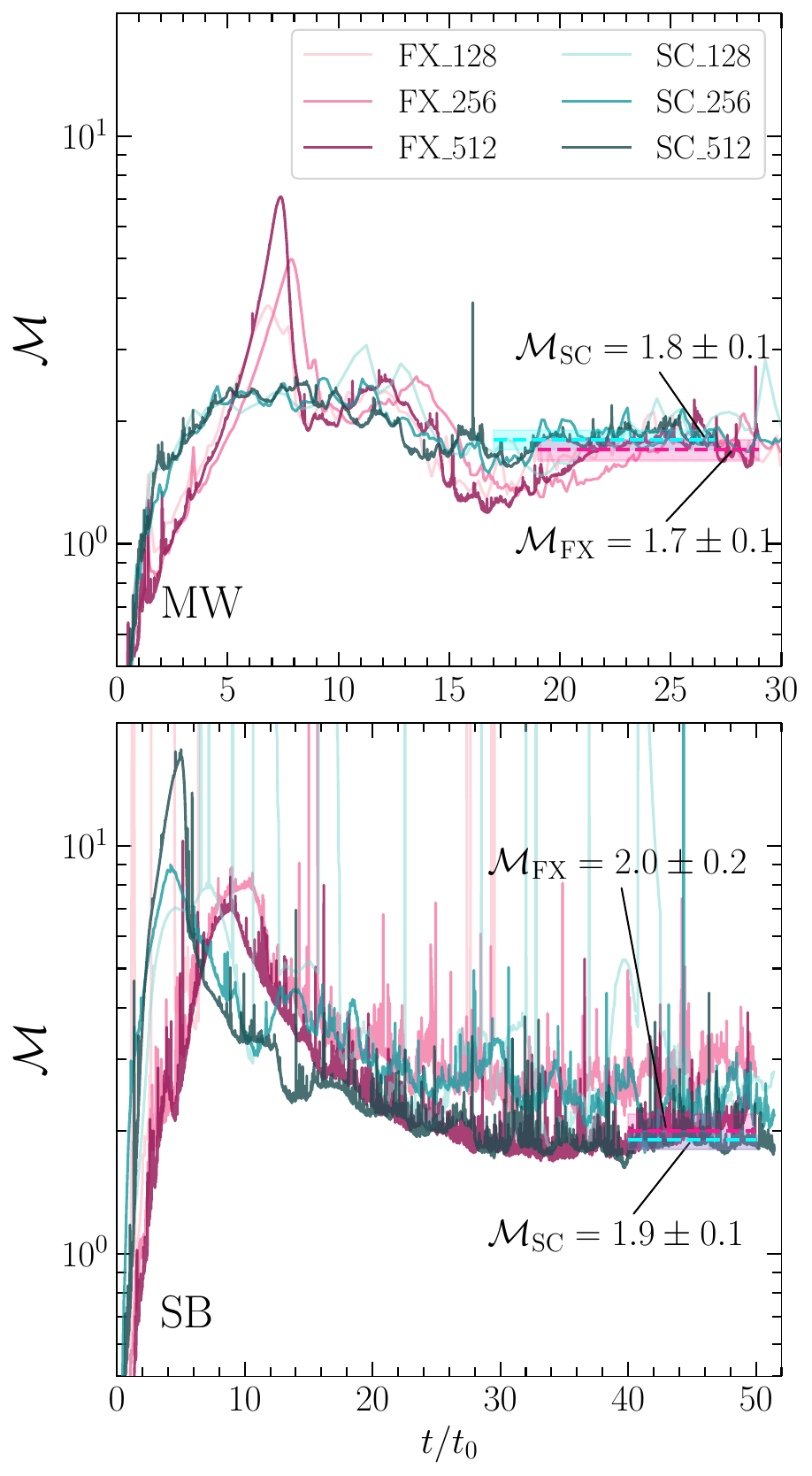} 
        \caption[]{
        The global, volume-weighted turbulent Mach number evolution, $\M$ (\autoref{eq:mach}) , as a function of time in units of the turbulent turnover time on the scale-height, $t/t_0$, for each simulation (shown in Table~\ref{tb:results}). SB simulations are shown in the top panel and MW in the bottom, with each color indicating a different grid resolution of the FX (pinks) and SC (blues) models. For the SB and MW models, the turbulence reaches an approximately statistically stationary state for $t \gtrsim40 t_0$ and $t \gtrsim 17-19t_0$, respectively. All quantities in this study will be averaged over $10t_0$ after the stationary state is reached, indicated by the dotted lines for each model. In general, all galaxy models are super-to-trans-sonic, meaning that the systems are dominated by velocity rather than thermal fluctuations. We find for MW, $\M_{\rm SC} \approx 1.8$ and $\M_{\rm FX} \approx 1.7$, aligning with observations of the WIM in the Milky Way \citep{Gaesnsler_2011_trans_ISM}. For SB we find $\M_{\rm SC} \approx 1.9$ and $\M_{\rm FX} \approx 2.0$, aligning with WIM observations of M82 \citep{Westmoquette2009_velocity_dispersion_M82}. The fluctuations in $\M$ (associated with intense spikes from SNe energy and momentum injections; \citealt{Kolborg2022_metal_mixing_1}) decrease with increased resolution, most likely due to the reduction of the volume-filling factor for the SNe seeds.}
        \label{fig:mach}
    \end{figure}  

\subsubsection{Self-consistent seeding (SC)}\label{sssec:sc}
\begin{figure*}[htbp]
        \centering
        \includegraphics[width=1\linewidth]{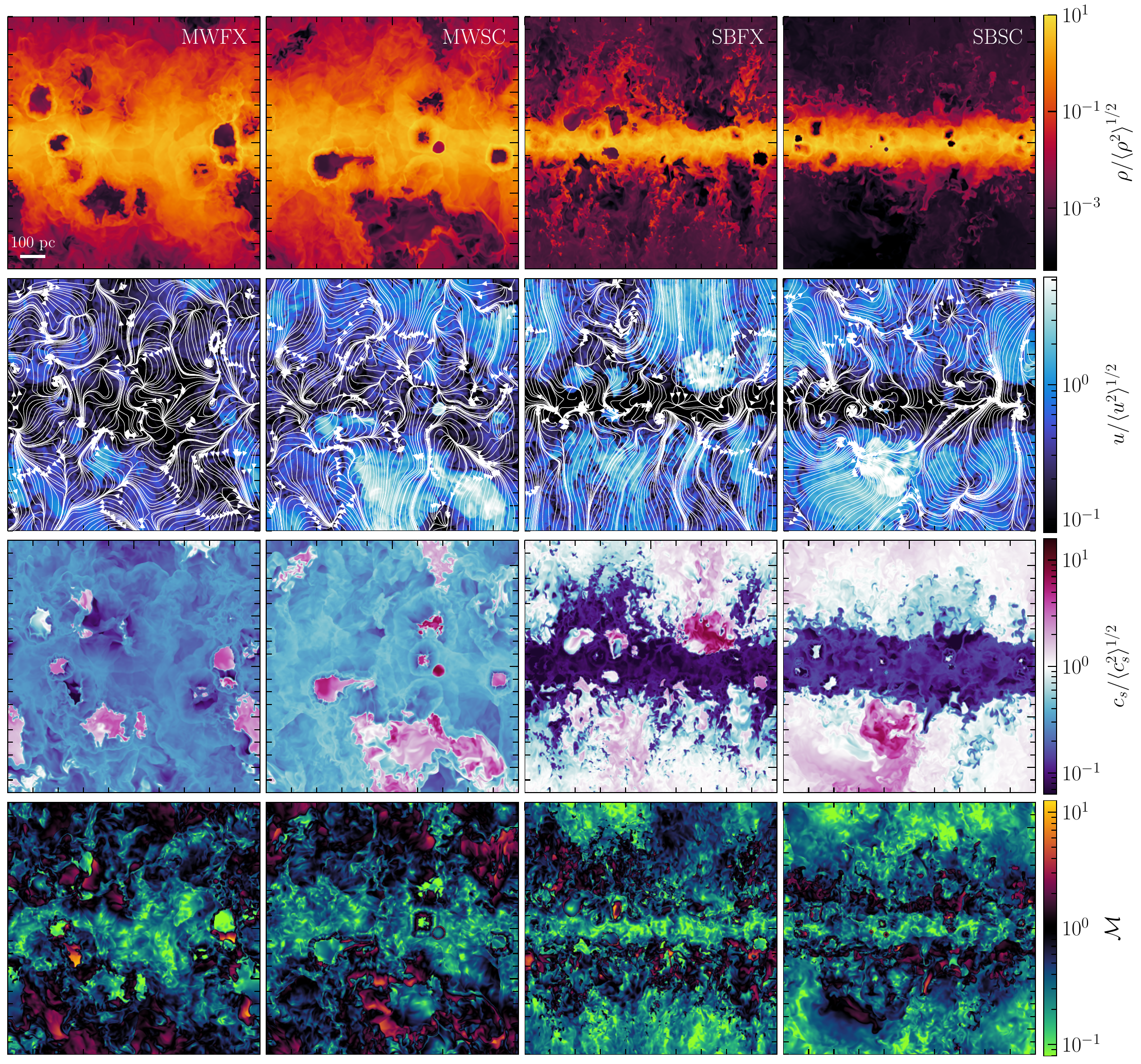}
        \caption{Two-dimensional slices parallel to $\bnab\phi$ of statistically stationary field quantities that are used throughout this study. The panels are organized by the different galactic models (MW, first two columns, and SB, last two columns) and seeding prescriptions (FX and SC suffixes), with annotations shown in the top-right of each panel in the first row. The top three quantities are normalized by their respective volume-weighted root-mean-squared value. The first row is the mass density, $\rho$, the second row is the velocity magnitude, $u$, with velocity streamlines in the slice plane shown in white, the third row is the thermal sound speed, $c_s \propto \sqrt{P/\rho}$, and the last row is the local Mach number, $\M = u/c_s$. The MW models (first two columns) are characterized by a thick, transonic, warm disk, and a SNe explosion rate of $\gamma_{\rm SNe}/10^{-4}\,\rm{yrs} = 0.1$. Whereas the SB models are characterized by a thin disk, hot, diffuse wind, and a $\gamma_{\rm SNe}$ an order of magnitude larger than the MW model.}
        \label{fig:slice}
    \end{figure*}

    We have a considerably more physically motivated approach with the SC seeding model, intended to mimic the way in which CC-SNe are preferentially clustered near the dense gas where massive stars tend to form and live \citep{Galbany_2018_SNE_dense,deWit_2004_OB_clustering, LADA_stars_in_dense}. In this scheme, the input global rate, $\gamma_{\rm{SNe}}$, shown in Table~\ref{tab:ics}, determines when the simulation enters into a SNe explosion loop. Once in the loop, we use the local $\rho$ and $\phi$ to determine if a SN detonates. Specifically, for each cell with volume $\mathcal{V}_{\rm{cell}}$, at each timestep $\Delta t$, a random sample is drawn from a Poisson distribution, $p(n|\lambda) = \lambda^k e^{-n}/n!$,  with $n$ explosion occurrences and mean, $\lambda$, where
    \begin{align}\label{distsc}
        \lambda(x,y,z) = \dot{n}_{\rm{SNe,SC}}\mathcal{V}_{\rm{cell}}\Delta t,
    \end{align}and\begin{align}\dot{n}_{\rm{SNe,SC}}=\epsilon_*\frac{\rho(x,y,z)}{t_{\rm{ff}}(z)100\rm{M}_{\odot}}\label{ratesc},
    \end{align}
    where $\dot{n}_{\rm{SNe,SC}}$ is the local volumetric SN rate, $\epsilon_*=1/100$ is the star formation efficiency, and $t_{\rm{ff}}(z)$ is the local free-fall time scale in $\bnab\phi$, which is found using the dynamical density, $\rho_{\rm ff}(z)$ in the stellar component of $\bnab\phi$,
    \begin{align}\label{t_ff}
        t_{\rm{ff}}(z)=\sqrt{\frac{3\pi}{16G\rho_{\rm ff} (z)}},\quad\text{where} \quad
        \rho_{\rm ff}(z)=\frac{\partial_z^2\phi(z)}{4\pi G}.
    \end{align}
    If $n > 0$, a single SN is seeded at that position. Because $\lambda \propto \rho/t_{\rm ff}$, the probability favors dense regions with short free-fall times, predominately near the midplane. Therefore, there is no strict $z$ threshold where SNe can explode, like in the FX prescription. Hence, there is a nonzero (but small, since $t_{\rm ff}(z)$ becomes longer and $\rho$, more dilute) probability for a SNe to be seeded outside of the midplane, in the outer winds of the galaxy, visible in figures \ref{fig:mask} and \ref{fig:slice}. These rare occurrences reflect runaway OB stars exploding in the more diffuse ISM \citep{Conroy_2012_runawayOB, Stone_1991_runawayOB}. 

\subsection{Statistical stationarity} \label{ssec:stationarity}
    Determining characteristic time scales of turbulence is critical to this analysis and turbulence studies in general, as these time scales dictate the duration over which the statistics are measured from the models. As the simulation begins and evolves, the gas cools and collapses in $\bm{\nabla}\phi$, while SNe are injecting mass, energy and momentum into the disk \citep{Kolborg2022_metal_mixing_1,Kolborg2023_metal_mixing_2}, in accordance to the seeding schemes. As the simulations progress, the SNe drive turbulence in the ISM through baroclinicity \citep{Beattie2025_so_long_k41} and contribute to pressure support of the gas, both thermally and non-thermally. The disk and outflow eventually reach a statistically steady-state where the flux from the SNe injections is equal to the dissipated flux \citep{Beattie2025_so_long_k41}, and a stationary turbulent cascade forms, becoming invariant over time. 
    
    Analogously with standard turbulence box studies \citep[e.g.,][]{Federrath2013_universality,Burkhart2020_CATS,Beattie2025_nature_astronomy}, we sample the turbulence across multiple turnover times, $t_0$, to ensure that our results are robust to fluctuations across multiple statistical realizations. We define $t_0$ for each model,
    \begin{align} \label{eq:t0}
        t_0=\frac{\ell_0}{\langle u^2 \rangle^{1/2}},
    \end{align}
    where $\ell_0$ is the scale height of the gaseous disk in steady state, detailed in \autoref{ssec:scaleheight}. Heuristically, this can be thought of as the time it takes for a turbulent fluctuation to transverse the scale height, but as demonstrated later in \autoref{ssec:universal_cascades} and in \citet{Beattie2025_so_long_k41}, it is not the outer scale of the turbulence cascade, which extends well into the winds. As discussed in \citet{Grete2025_density_distribution}, it is not obvious that we can associate this with the time it takes the turbulence to decorrelate on this scale (Strouhal number $=1$). 
    
    To identify where statistical stationarity begins for each model, we first calculate the volume-integral turbulent Mach number, 
    \begin{align}\label{eq:mach}
    \M = \left\langle \left( \frac{u}{c_s} \right)^2 \right\rangle^{1/2},
    \end{align}
    where $u$ is the local, non-thermal, rest-frame velocity, and $c_s$ is the thermal velocity. This is using the same definition as \citet{Fielding2022_ISM_plasmoids} and \citet{Beattie2025_so_long_k41}, where $c_s = \sqrt{\gamma P/\rho}$ is calculated locally and included in the volume-integral, which then includes the covariance terms between the temperature and the fluid velocity. We plot the evolution of $\M$ for each model in \autoref{fig:mach} as a function of $t/t_0$, where $t$ is the simulation time, and identify a range where the time-averaged $\M$ no longer significantly changes for different $\Delta t$ time windows, $\langle \M(t)-\M(t+\Delta t)\rangle_{\Delta t}\approx0$. This steady-state period begins at approximately $t\gtrsim40t_0$ for SB and $t\gtrsim17t_0$ and $t\gtrsim19t_0$ for MWSC and MWFX, respectively, denoted by the dashed lines in \autoref{fig:mach}. In this study, we perform all of our analysis averaged over $10t_0$ after steady state begins, unless explicitly stated otherwise. 

\section{ISM diversity}\label{sec:ISM_diversity}
    In \autoref{fig:slice} we show two-dimensional slices through the center of the domain, parallel to $\bnab\phi$, for each of the four galaxy and seeding models (MW and different seeding schemes in the left two columns, and SB and different seeding schemes in the right two columns). We normalize all field variables by their volume-weighted root-mean-squared value. The first row is the mass density, $\rho/\left\langle \rho^2\right\rangle^{1/2}$, the second row is the velocity magnitude $u/\left\langle u^2\right\rangle^{1/2}$ with overlaid velocity streamlines, the third row is the thermal sound speed, $c_s/\left\langle c_s^2\right\rangle^{1/2}$ and the final is the local $\M = u/c_s$ (for which we take the rest-frame root-mean-squared to calculate the turbulent $\M$ in \autoref{fig:mach}). 

   The differences between the MW and SB models are apparent in \autoref{fig:slice}, which will be a recurring theme in our study. The MW models feature thick, transonic, warm disks that extend throughout most of the vertical domain. In contrast, the SB models have thin, dense disks characterized by hot, subsonic winds that occupy a significant portion of the volume. This distinction is especially evident in the velocity streamlines shown in SBFX. The SNe seeding difference is particularly noticeable in the SB simulations.  The FX runs show filamentary structure due to many SNe going off at the disk boundary, but none above $80$pc, ejecting dense material into the winds which cools and falls back to the disk. In the SC runs, many SNe are heating and puffing up the disk, but undergoing radiative losses in this region contributing to a less-developed wind. Overall, this dichotomy between hot, wind-dominated volume-integral statistics and warm, disk-dominated statistics between the two models makes for a robust study across two very qualitatively different ISMs.  

\subsection{Turbulent Mach number} 
\label{ssec:mach_disc}
    Along with determining the characteristic timescales in our simulations in \autoref{ssec:stationarity}, \autoref{fig:mach} confirms that the volume-integral properties of the turbulence in these simulations align with approximately observations. Volume-averaged over the entire ISM, we find $\M_{\rm{SC}}=1.8 \pm 0.1$ and $\M_{\rm{FX}}=1.7\pm0.1$ for the MW model (consistent within $1\sigma$), in agreement with radio observations of the transonic WIM in the Milky Way \citep[$\M\approx2$;][]{Gaensler2011}. 
    
    For our SB model, $\M$ can intermittently reach $\M\gtrsim10$. High $\M$ ISMs are typical of starburst galaxies (see $\M\approx96$ for cold star-forming clumps in starburst M82, \citealt{2018MNRAS.477.4380S}). Our model is on the low-SFR end of starbursts, more analogous to a metal-poor starburst galaxy \citep{Kennicutt_MW_SB}. Regardless, we find $\M_{\rm{SC}}=1.9\pm0.1$ and $\M_{\rm{FX}}=2.0\pm0.2$ for the volume-filling component of the ISM, comparable to our MW model and starburst galaxy ionized gas observations. Using non-thermal line broadening from the ionized gas in M82, $\left\langle u^2\right\rangle^{1/2} \approx 23-95\,\rm{km\,s}^{-1}$ \citep[where we have included a factor $\sqrt{3}$, assuming isotropy, and converted the FWHM measurements into $\left\langle u^2\right\rangle^{1/2} = \sigma_u$;][]{Westmoquette2009_velocity_dispersion_M82}, and assuming a standard WIM thermal velocity $c_s \approx 10\,\rm{km\,s}^{-1}$, this gives $\M \approx 2 - 10$ in the WIM for M82, with a lower bound that is broadly consistent with our simulations. This shows, regardless of both the $\gamma_{\rm SNe}$ and the parameterization of the potential, the volume-integral $\M$ remains relatively constant across the galaxies, even when the $\gamma_{\rm SNe}$ increases by an order of magnitude, indicative of a strong covariance in $c_s$ and $u$ fluctuations.

    As noted in \citet{Kolborg2022_metal_mixing_1,Kolborg2023_metal_mixing_2} and above, individual SN explosions can appear as visible spikes in $\M$. However, we note that these fluctuations decrease drastically in intensity from $N_{\rm{grid}}=128$ to $N_{\rm{grid}}=512$. This is likely a numerical effect, due to the reduction of the volume-filling factor for the SN seeds, as injections move to progressively smaller scales with increased resolution (see \autoref{ssec:Seeding}), e.g., the SN explosions takes up a significant portion of the disk at low $N_{\rm{grid}}$, whereas the resolved WIM takes up more of the volume as $N_{\rm{grid}}$ increases. This effect is especially strong for the SB model, which we show in \autoref{fig:slice} \& \autoref{ssec:scaleheight} has a much thinner, more compact disk, and also has a higher global $\gamma_{\rm SNe}$ (Table~\ref{tab:ics}), together leading to a well-developed wind outflow.

\subsection{Scale height \& galaxy profiles}\label{ssec:scaleheight}

\begin{figure}[htbp]
        \centering
        \includegraphics[width=0.85\linewidth] {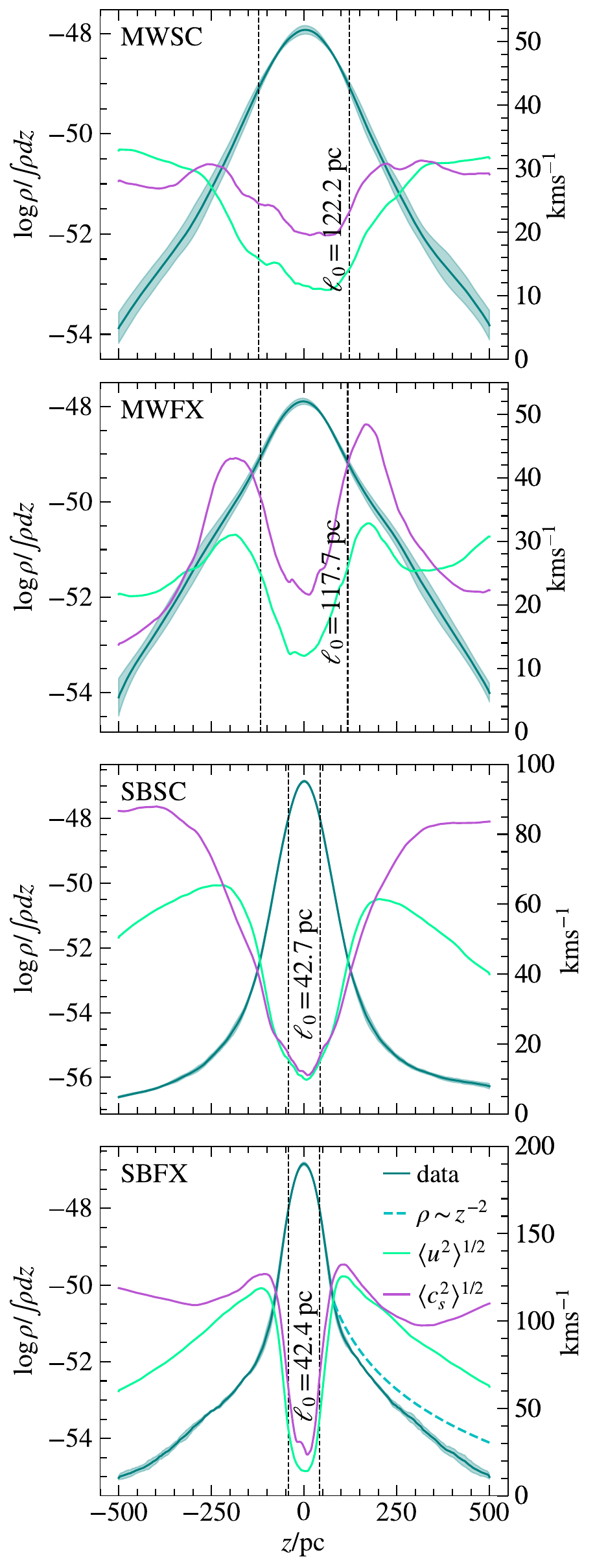}
        \caption[]{
        Time-averaged vertical profiles for each model (top-left corner annotation). We show the log-density profile normalized by the total column density (teal, left axis), the velocity rms $\left\langle u^2\right\rangle^{1/2}_{x,y}$ (green, right axis), and the sound speed rms $\left\langle c_{s}^2 \right\rangle_{x,y}^{1/2}$ (purple, right axis). The gas scale height $\ell_0$ is found by fitting \autoref{eq:isothermal_profile} and using \autoref{eq:scaleheight}, and is annotated with the black, dashed lines in each panel. It varies between $\ell_0 \approx 40-120\,\rm{pc}$ between the SB and MW models, respectively. For the SBFX model (bottom), we overlay the mass density profile solution for adiabatic winds, $\rho \propto z^{-2}$ \citep{Chevalier_1985_Nature}. 
        }
        \label{fig:4panel_density}
    \end{figure}

    The necessary condition for calculating the characteristic timescale, $t_0$ (\autoref{eq:t0}), is the definition of a characteristic length scale. We measure the scale height of the gaseous disk in steady state after evolution from initial conditions, as opposed to the aforementioned $z_{\rm{eff}}$, which is an analytical scale height derived from initial conditions to determine $z_{\rm{SNe}}$ for the FX scheme in \autoref{sssec:fx}, not including any SNe feedback. We do so by fitting an analytical model, detailed below, to the time-averaged density profile of each simulation in \autoref{fig:4panel_density}. 

    In steady-state pressure equilibrium, \autoref{eq:momentum_conservation} becomes\begin{align}\label{eq:pressure_equilibrium}
        \frac{dP}{dz} = -\rho(z) \frac{d\phi}{dz},
    \end{align}
    where we have neglected the SNe driving term for simplicity. Consider an isothermal atmosphere,
    \begin{align}
        P(z) = \sigma_z^2\rho(z),
    \end{align}
    where 
    \begin{align} \label{eq:sigma_z}
        \sigma_z^2\approx \left\langle c_s^2 \right\rangle_{x,y} + \left\langle u^2\right\rangle_{x,y}
    \end{align}
    contains both thermal and turbulent pressure components. For constant $\sigma_z^2$, \autoref{eq:pressure_equilibrium} is separable, with solution,
    \begin{align}\label{eq:isothermal_profile}
        \frac{\rho(z)}{\rho_0 } = \exp\left[ -\frac{2\pi G}{\sigma_z^2} \left(
\Sigma_* \left( \sqrt{z^2 + z_0^2} - z_0 \right) +
 \frac{\rho_{\text{halo}}}{3} z^2
\right) \right].
    \end{align}
    We fit $\rho(z)$ to the density profile using maximum likelihood fitting, constructing the posterior by utilizing \textsc{emcee} \citep{emcee}. Fits are shown in \autoref{fig:4panel_density_app}.
     We define the gaseous scale height $\ell_0$ as the height $e$-folding distance from the midplane value,
    \begin{align}\label{eq:scaleheight}
        \ell_0 \approx \sqrt{\left(\frac{\sigma_z^2}{2\pi G \Sigma_*} + z_0 \right)^2 - z_0^2}.
    \end{align}
    inside the stellar component of the potential. This is shown in black dashed lines for each simulation. 

    We find in \autoref{fig:4panel_density} that $\ell_0$ is similar across seeding schemes for each model, with $\ell_{0,\rm{SC}}\approx126\,\rm{pc}$ and $\ell_{0,\rm{FX}}\approx121\,\rm{pc}$ for the MW model \citep[in agreement with other works for the vertical gaseous disk, see][]{Vijayakumar_2025}. For SB, we find  $\ell_{0,\rm{SC}}\approx43\,\rm{pc}$ and $\ell_{0,\rm{FX}}\approx44\,\rm{pc}$, which is in agreement with observational constraints  derived for   gaseous disks in  starbursting regions \citep[e.g.,][]{elmegreen_2025_scaleheight_SB}, and radio observations of the starburst galaxy M82, \citep[$\approx20\rm{pc}$;]{adebahr_m82_scaleheight}. The disk is slightly thicker for the SC runs, because more SNe are seeded compactly in the midplane, contributing to pressure support and expanding the disk. As expected, the disk of the SB galaxy is considerably thinner than that of MW and incredibly wind-dominated, due to the deeper gravitational potential and higher $\gamma_{\rm{SNe}}$. This is also shown by the order-of-magnitude difference in velocity and sound speed fluctuations between MW and SB, plotted in green and purple, demonstrating that the SB host a turbulent, hot wind. 

    The velocity dispersions predicted by both SB seeding models are consistent with observational findings for the ionized gas in starburst galaxy M82, as reported by \citet{Westmoquette2009_velocity_dispersion_M82} and discussed in \autoref{ssec:mach_disc}. The velocity dispersions for the MW models are $\approx 10\,\rm{km\,s}^{-1}$ in the disk, and reach $\approx32\,\rm{km\,s}^{-1}$. This aligns with the observational findings of ionized gas in the inner $50-100\rm{pc}$ of our Galaxy when results are transformed into 3D volume-weighted values \citep[$7-14\,\rm{km\,s}^{-1}$]{Langer_DIG}, and is also also consistent with previous expectations and simulation results \citep{Ferriere2020_reynolds_numbers_for_ism,Gent_2017}. The FX runs overall have larger thermal fluctuations and a more dynamic wind, because a high fraction of SNe are exploding in low-density regions (winds and disk boundaries), efficiently heating the gas and driving outflow. This is in contrast to the SC runs, where many SNe are seeded in the dense midplane, undergoing significant radiative losses before energy can contribute efficiently to the winds. Furthermore, it should be noted that for both FX runs, there is a visible negative turning point in the slope of $\langle u^2\rangle^{1/2}$ and $\langle c_s\rangle^{1/2}$, which is $z \approx z_{\rm{SNe}}$ and SNe are no longer seeded. 
    
    As mentioned in \autoref{ssec:mach_disc}, there exists a strong covariance between $\langle u^2\rangle^{1/2}$ and $\langle c_s^2\rangle^{1/2}$, resulting in a mostly transsonic $\M \approx 0.5-2$ throughout the $z$ profile. For SBFX, we overlay for comparison the \citet{Chevalier_1985_Nature} solution for adiabatic starburst galaxy winds (turquoise dashed), which yields a scaling of $\rho\propto z^{-2}$ at $z>R$, where R is the region of mass and energy production, which we set to $Z_{\rm{SNe}}$. They assume a perfectly adiabatic, spherically symmetric flow with no external potential and constant velocity, differing from our model, which may explain the slight deviation, likely most affected by $\bnab\phi$ and non-adiabatic processes (see \autoref{fig:phase}; SBFX).

\subsection{Mass density distribution functions}\label{PDFs}
\begin{figure}[htbp]
        \centering
        \includegraphics[width=0.9\linewidth] {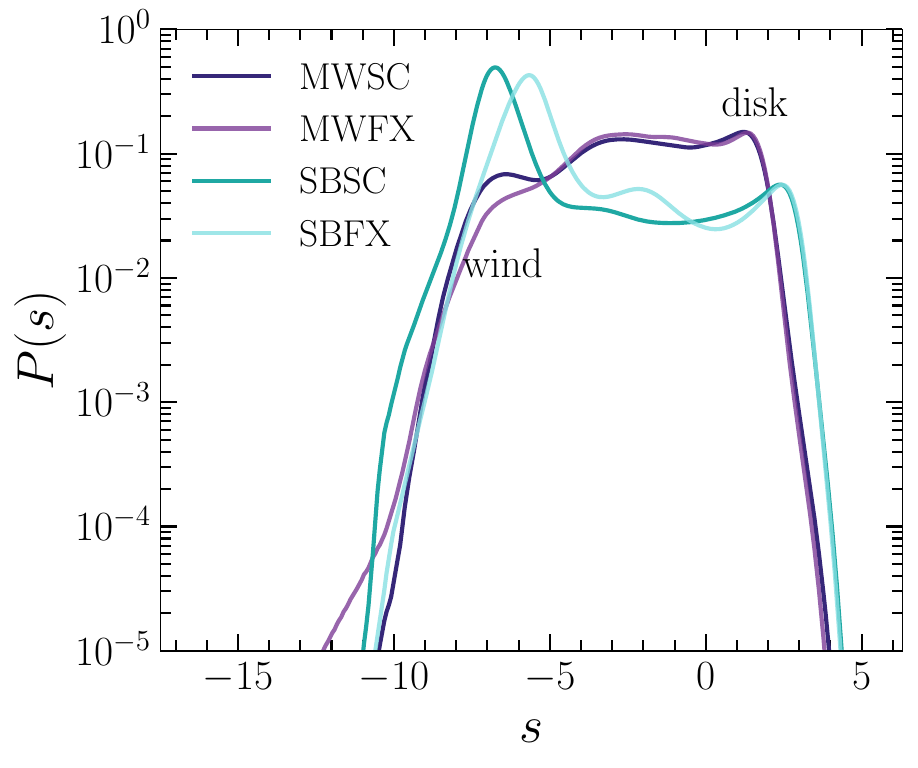}
        \caption[]{
        Time-averaged, one-dimensional, volume-weighted probability density functions of the logarithmic density contrast,  $s=\ln\left(\rho/\langle\rho\rangle\right)$, for MW (purples) and SB (blues) simulation models. We annotate the regions roughly corresponding to the disk and winds, demonstrating the contrast of the density distributions between galactic models. SB showcases bimodality in probability density from the winds and disk, with most of the density in the winds, and MW comprises predominantly of dense disk. 
        }
        \label{fig:density_PDF}
    \end{figure}

    In order to further analyze the density state in the ISM for each of our simulations, we plot the volume-weighted PDFs of the logarithmic density contrast, $\s=\ln(\rho/\langle\rho\rangle)$ in \autoref{fig:density_PDF} for all of the material in steady state. The two-dimensional $P-\rho$ and $T-\rho$ PDFs are also shown in \autoref{fig:phase}, which will be discussed in the following section. \autoref{fig:density_PDF} captures the prominence of over-dense and under-dense mass density statistics for each model, with a strongly visible wind and disk component. 
    
    For the SB model, the systems have prominent bimodality from the wind (at lower $s$) and disk (at higher $s$). The volume-filling factor is significantly larger for the diffuse winds. For the MW model, it is visible that the simulations are strongly disk-dominated with a weakly developed wind, and due to our initial conditions, the peak $s$ is lower than that of the SB model. Overall, it is well-known that the volume-weighted $s$-PDF for strongly-shocked, high-$\M$ turbulence has a negative skewness and truncated high-$s$ tail,  due to the low volume-filling factor of dense filamentary and shocked material which comprise this portion of the PDF \citep{Burkhart2009_bispectrum,Hopkins2013_non_lognormal_s_pdf,Federrath2008,Mocz2019,Beattie2022_spdf}. For testing these models, we would have to do a detailed decomposition of $s$ into phases, as in \citet{Kim2023_multiphase_ISM}. We do not do the phase decomposition in this study, and the main takeaway is that the two models have significantly different $s$ (and hence mass density) statistics. 

     \begin{figure}[htbp]
        \centering
        \includegraphics[width=0.9\linewidth] {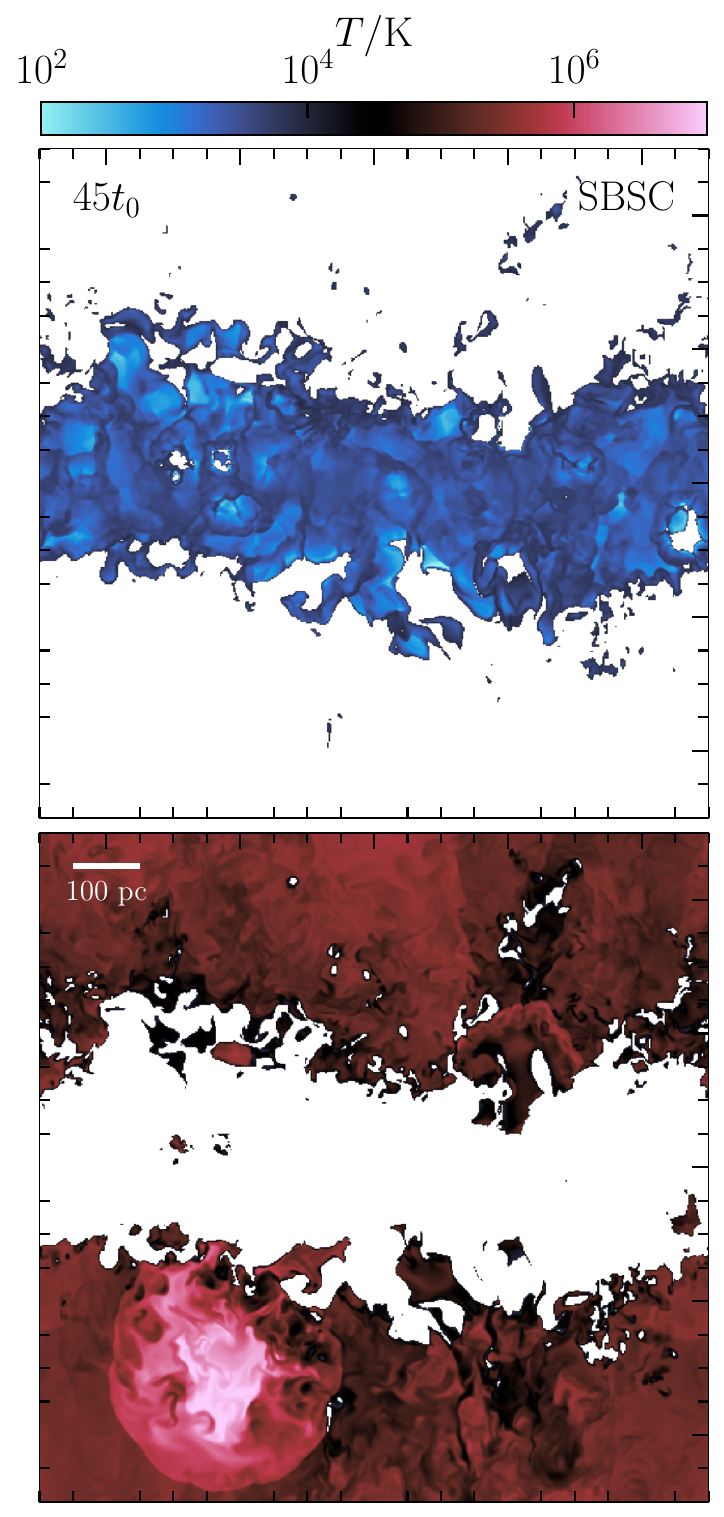}
        \caption[]{Two-dimensional ($1\,\rm{kpc}^2$) temperature slices along the direction of $\bnab\phi$ for the SBSC simulation in steady state. We mask the temperatures to show material at $T<10^4\,\rm{K}$ (top panel) and $T>10^4\,\rm{K}$ (bottom panel). A hot ($T > 10^6\,\rm{K})$, expanding SN remnant is visible in the bottom left corner. This verifies that the disk primarily resides at $T \approx10^3\,\rm{K}$, and the winds mostly $T\gtrsim10^4\,\rm{K}$.}
        \label{fig:mask}
    \end{figure}

    \begin{figure*}[htbp]
        \centering
        \includegraphics[width=0.92\linewidth]{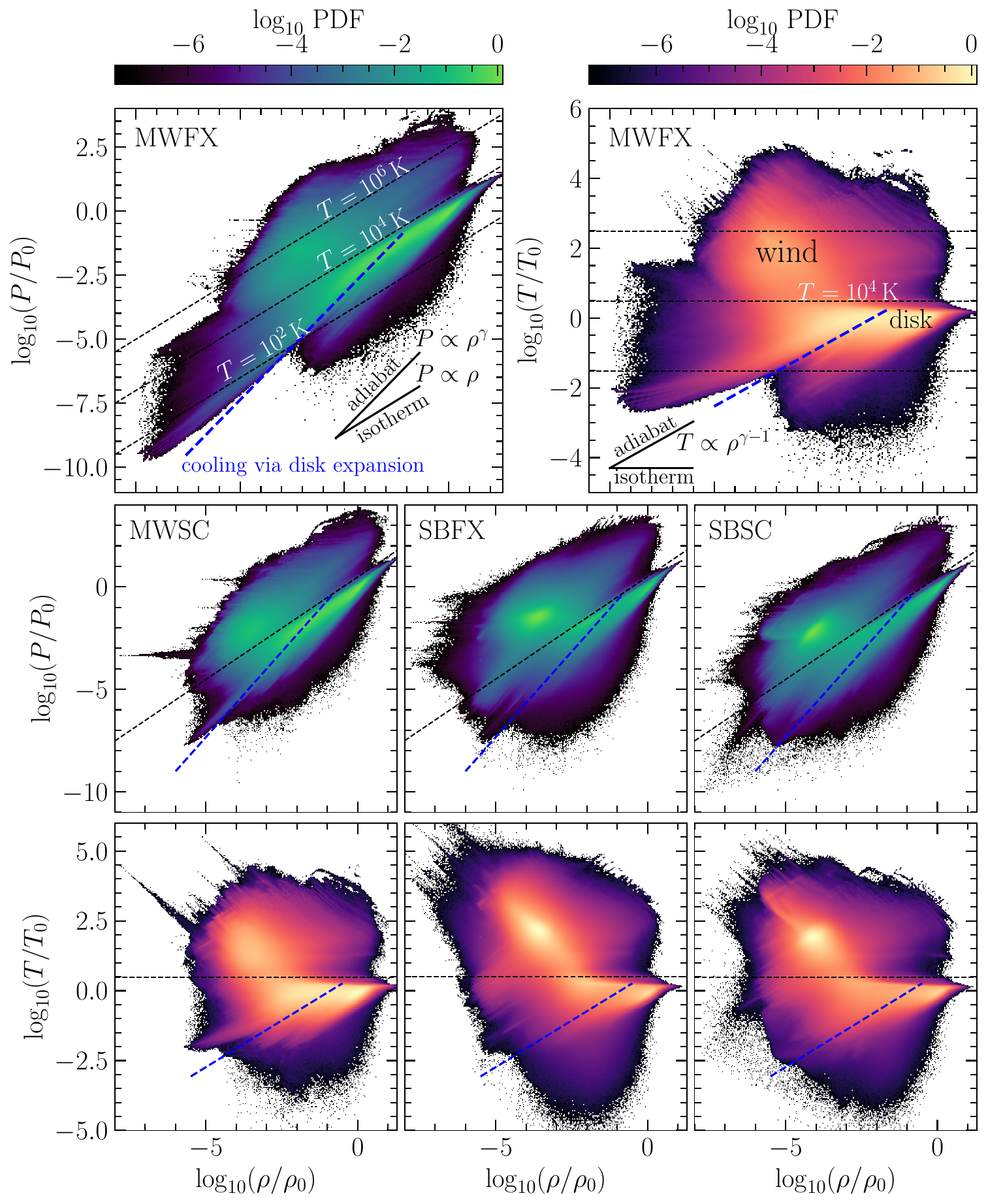}
        \caption[]{Time-averaged, two-dimensional pressure-density (top left, middle row) and temperature-density (top right, bottom row) probability distribution functions (or phase plots) for each model, normalized by average midplane values in steady state. Our fiducial MWFX model is enlarged on the top row as a guide to key features. This illustrates the presence of a multiphase ISM with a volume-filling warm phase of $T\lesssim 10^4\,\rm{K}$ (WNM and WIM) and a hot $T\approx10^6\,\rm{K}$ (HIM) phase. $\rho$ and $P$ have been normalized by their average midplane values, presented in Table \ref{tb:results}. In the top panels, example adiabats $P\propto \rho^{\gamma}$ and isotherms $P\propto\rho$ are plotted, and the hot wind region (above $10^4\,\rm{K}$) and disk midplane (below $10^4\,\rm{K}$) are highlighted on the right. All cooling to $T \lesssim 10^4\,\rm{K}$ is achieved via adiabatic cooling, illustrated by the example blue, dashed adiabat, via expansion of the galaxy disk and SNR in the disk. The distributions show the varied phase structure of the ISM across seeding schemes and galaxy models. }
        \label{fig:phase}
    \end{figure*}

\subsection{Interstellar medium phases}\label{ssec:phase}
    To better understand the relationship between our cooling function, the density structure, and the ISM phases in our simulations, as well as further compare the ISM properties of our models, we plot the time-averaged phase diagrams of $P$ and $T$ as a function of $\rho$, all normalized by their respective average values at the midplane ($z=0$; see Table~\ref{tb:results}). Isotherms are plotted at $T=10^2\,\rm{K}$, $10^4\,\rm{K}$, and $10^6\,\rm{K}$, and as mentioned in \autoref{ssec:cooling}, the volume-filling WNM, WIM, and HIM phases are the focus of our study. 

    For all models, there is a clear build-up of flat probability density below the $T=10^4\,\rm{K}$ isotherm around the disk midplane which lies (on average) at $T=3,\!000$-$3,\!500\,\rm{K}$ (see Table~\ref{tb:results}). The region above $T=10^4\,\rm{K}$ is primarily the warm-hot wind and outflow region from SNe feedback. We have plotted an example temperature map for the SBSC model in \autoref{fig:mask}, masked to show material at $T>10^4\,\rm{K}$ (primarily winds) and $T<10^4\,\rm{K}$ (mostly disk), to verify the strong temperature boundaries present between the winds and disk. As mentioned in \autoref{ssec:cooling}, our cooling function mostly truncates at $T=10^4\,\rm{K}$, but material can still cool below this from other mechanisms, most efficiently via adiabatic expansion of the galaxy disc and SNe therein (see \autoref{eq:cs_uc}). In other words, nearly all the material below $T=10^4\,\rm{K}$ follows an adiabat, and almost all this material is in the disk. This adiabatic trajectory, $P\propto\rho^{\gamma}$ (where $\gamma=5/3$ for monoatomic gas), is most directly visible in the phase space paths from $T=10^4\,\rm{K}$ to $T=10^2\,\rm{K}$, one of which is plotted as a blue dashed line on the top-left plot. Any deviation from these adiabatic trajectories is likely due to the non-zero cooling function having small effects in the lower temperature phases, detailed in \autoref{ssec:cooling}. We also see faint adiabatic cooling tracks in the WIM and HIM, due to the expansion of SNR in the diffuse winds. 

    As mentioned previously, the disk of the MW model has a much higher volume-filling factor than the SB model, due to lower $\rho_0$ and shallower $\bnab\phi$. Alternatively, the SB model has a higher volume-filling factor in the winds, with a significantly higher $\gamma_{\rm SNe}$and deeper $\bnab\phi$, developing significant outflows. For both the SC simulations, and particularly for the SBSC, simulation, we see a larger and smoother spread of phase-space in the disk region and below, and therefore a higher volume of gas cooling to below $T\approx10^3\,\rm{K}$ compared to SBFX, likely due to the fact that there are more SNe detonating in the dense midplane in this simulation, which increases the presence of adiabatic expansion of SNe (see \autoref{fig:mask}). Furthermore, for the SBSC simulation (right, second and third row), we see a buildup of probability density on an adiabatic trajectory above $T=10^4\,\rm{K}$, just between $\log_{10}P/P_0=0$ and $\log_{10}P/P_0=-5$. This is due to the fact that we have SNe further into the winds than is prescribed in SBFX (see the location of the SNR in \autoref{fig:mask}), so the SNR expand with less ambient pressure confinement and surrounding material, persisting in the adiabatic Sedov-Taylor evolution for longer. 
    
    Faint isobaric tracks can be observed in all phase diagrams, which follow the pressure-driven snowplow radiative cooling phase of SNe expansion, but there is a strong isobaric signature above $T=10^4\,\rm{K}$ in the SBFX plot (middle panel). Here, many SNe are exploding in lower wind regions, meaning they are interacting with denser, higher pressure, cooler environments, shortening the adiabatic Sedov-Taylor phase and spending more time in the snowplow radiative cooling phase. Overall, the SB model shows stronger signatures related to SNe phases, due to the much higher SNe rate compared to the MW model. 

    The important takeaway of the ISM analysis we have done throughout this section is the contrast in dynamical, mass density, and thermal structure between models, revealing how different supernova driving conditions and galactic models create markedly different interstellar phase environments. 

    \begin{figure*}[htbp]
        \centering
        \includegraphics[width=0.9\linewidth]{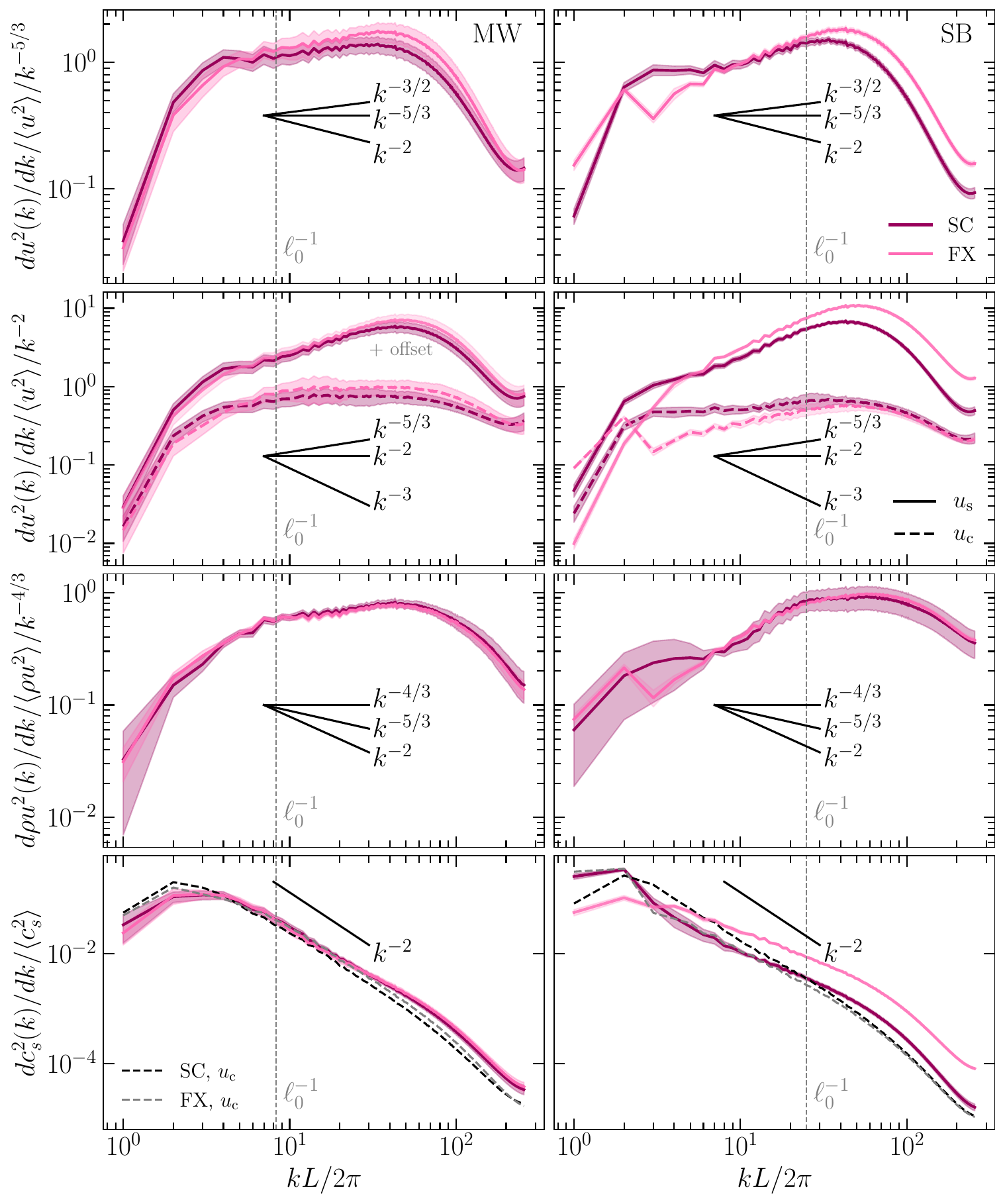}
        \caption[]{
        One-dimensional, isotropic power spectra normalized by the integrated power, as a function of wavenumber, $k$, normalized by the system scale wavenumber, $2\pi/L$, with the gaseous scale height, $k_0 \sim \ell_0^{-1}$, annotated on each panel with the grey vertical line. The shaded regions show the $1\sigma$ fluctuations of the spectra in time. We show MW (left column) and SB (right column), for both seeding mechanisms (dark and light colors). All fitted indexes for spectra are shown in Table~\ref{tb:slopes}. \textbf{Row one:} the velocity spectrum compensated by $k^{-5/3}$ for \citet{Kolmogorov1941}-type turbulence. \textbf{Row two:} the velocity spectrum decomposed into compressible $\u_c$ (dashed) and incompressible $\u_s$ (solid) modes for each seeding mechanism (offset for clarity), compensated by $k^{-2}$ \citet{Burgers1948}-type turbulence, and with annotations for the quasi-acoustic turbulence $k^{-3}$ \citep{Wang_2017_3_2,Beattie2025_compressible_dynamo}.
        \textbf{Row three:} the kinetic energy spectrum compensated by $k^{-4/3}$ (found previously in compressible turbulence \citealt{Grete2021_as_a_matter_of_tension}). \textbf{Row four:} the $c_s$ spectrum, with $\u_c$ velocity spectra modes for each simulation plotted in dashed black and grey lines, showing that both spectra trace one another other (a natural repercussion of being in the adiabatic limit and mediated by acoustic waves, \autoref{ssec:cs_discussion}), following $\sim k^{-2}$ for the MW simulations. We show further relevant $k$-scaling annotations in black for each panel, which we discuss in the main text. 
        }
        \label{fig:powerspec}
    \end{figure*}
    
\section{Power spectra and turbulence in the galaxy models}\label{sec:power_spectrum}

     We have now demonstrated the clear differences between the ISM structures across each galaxy type and seeding prescription. In this section, we investigate the nature of the Fourier power spectrum of the turbulent gas. 

     \subsection{Definitions}
     Essential to our analysis of the turbulent structure in our simulations is the nature of the velocity modes, which originate from distinct physical processes in a galaxy \citep{Sharda2021_driving_mode,Dhawalikar2022_shock_driving_parameter,Menon2020b}. We decompose the velocity $\u$ into compressible $\u_c$ ($|\bnab \times \u_c| = 0$) and incompressible $\u_s$ ($\bnab \cdot \u_s=0$) modes, using a Helmholtz decomposition of the field, such that,
     \begin{align}\label{eq:H_decomp}
        \u = \u_c + \u_s, \quad \u_c \cdot \u_s = 0.
    \end{align}
    We do this in Fourier space, where 
    \begin{align}
        \Tilde{\u}_c(\k)= \frac{\k\cdot\Tilde{\u}(\k)}{k^2}\k, \quad \Tilde{\u}_s(\k) = \Tilde{\u}(\k) - \Tilde{\u}_c(\k),
    \end{align}
    and the Fourier transformed velocity is 
    \begin{align}
        \Tilde{\u}(\k) \propto \int \u(\bell)\exp\left\{-2\pi i \k\cdot\bell \right\} \d{\bell}. \label{eq:Fourier}
    \end{align}
    Then we simply take the inverse Fourier transforms of $\u_c$ and $\u_s$ to get the real-space velocities, which we can manipulate in the regular manner. Our implementation preserves both relations in \autoref{eq:H_decomp} to machine precision. 
    
    From a Fourier transformed field variable we can calculate the Fourier power spectrum, which we can use to probe the scale-dependent turbulence statistics and test turbulence models. It is
    \begin{align}
        \frac{df^2(\k)}{d\k} = \Tilde{\bm{f}}(\k)\cdot\Tilde{\bm{f}}^{\dagger}(\k),
    \end{align}
    where $\square^{\dagger}$ denotes the complex conjugate. Even though there is a global anisotropy imparted on the flow due to $\bnab \phi$, it is never strong enough to make the $\ell \parallel \bnab \phi$ and $\ell \perp \bnab \phi$ statistics significantly different \citep{Beattie2025_so_long_k41}. Hence, in this study we integrate the three-dimensional spectrum isotropically over solid-angle shells  $\d{\Omega_k}$, where $k=|\k|$, 
    \begin{align}
       \frac{ df^2(k)}{dk} = \int \d{\Omega_k} \; k^2 \frac{df^2(\k)}{d\k},
    \end{align}
    resulting in the one-dimensional isotropic spectrum, $df^2(k)/dk$.

    \subsection{The $k$-space structure of the turbulence}\label{ssec:power_spectrum}

    In \autoref{fig:powerspec}, we plot the integrated spectra of the total velocity, $du^2(k)/dk$ (top row), the kinetic energy $d\rho u^2(k)/dk$ (second row), the decomposed compressible $du_c^2(k)/dk$ and incompressible $du_s^2(k)/dk$ velocity modes (third row), and the sound speed $dc_s^2(k)/dk$ (bottom row), for MW (left column) and SB (right column), all normalized by the integral power. The gaseous scale height, $k_0 \sim \ell_0^{-1}$, is indicated with the grey vertical line. All fitted indexes for spectra are shown in Table~\ref{tb:slopes}, and the exact empirical indexes, with the caveats associated with measuring over a fixed range of $k$ in simulations that are not scale-free, are discussed in \autoref{app:slopes}.
    
    Beginning with the first row, we compensate the velocity spectra for \citet{Kolmogorov1941}-type turbulence, and we find that all models have a shallower than the \citet{Kolmogorov1941} spectrum, and MW models follow the $k^{-3/2}$ spectrum found in the magnetohydrodynamic (MHD) SNe-driven turbulence study of \citet{Gent2021_supernova_turbulence_and_dynamo} and also the MHD studies by \citet{Beattie2025_nature_astronomy} and \citet{Iroshnikov_1965_IK_turb}. This spectrum has also been derived for weak acoustic wave turbulence \citep{Zakharov_acoustic, Kochurin_2024_acoustic}, and shown to potentially be embedded within supersonic turbulence if one completely resolves a subsonic cascade \citep{Ferrand_2020,Beattie2025_nature_astronomy}. This is plausible for our set up if SNe shocks become linear sound waves on small enough scales.
    
    Once we decompose each simulation into $\u_c$ and $\u_s$ velocity modes (second row), we observe that, as found in a number of SNe-driven ISM simulations, e.g., \citet{Balsara2004_SNe_turbulence_and_dynamo}, \citet{Padoan2016_supernova_driving}, \citet{Luibun2016_SNe_driving_modes} and  \citet{Beattie2025_so_long_k41}, the turbulence is strongly dominated by the incompressible $\u_s$ modes, with a spectra shallower than $k^{-5/3}$, and especially so for the high-SN rate SB model. This aligns also with work that found incompressible mode dominance to be true even for purely compressible-driven simulations \citep{Federrath2013_universality}, but likely the mechanism for generating the incompressible modes varies significantly between the simulations, as discussed in \citet{Beattie2025_so_long_k41}, i.e., the shallower spectrum dominates at smaller scales than the scale height, even more so than in purely incompressible-driven turbulent boxes \citep{Federrath2013_universality}. \citet{Beattie2025_so_long_k41} proposed that phase mixing through the fractal cooling layer of SNRs is the strongest source of incompressible turbulence in SNe-driven turbulence. 
    
    The compressible mode spectrum in the second row closely follows \citet{Burgers1948} $k^{-2}$ prediction for strongly nonlinear compressible turbulence for the models, with some shallowing towards $k^{-5/3}$ in the SBFX model, but generally, $k^{-2}$ without a break scale. We also plot the pseudo-acoustic wave spectrum $k^{-3}$, which has been realized in low-$\M$ ($\M \lesssim 0.2$) hydrodynamical turbulence \citep{Wang_2017_3_2,Beattie2025_compressible_dynamo}, but as we see here, none of the simulations are in this regime.
    
    Because the turbulence is dominated by $\u_s$ modes it is unlikely that the $k^{-3/2}$ $\u$ spectrum comes from the weak acoustic regime, and in order to get shallower spectra analogous to MHD turbulence (like $k^{-3/2}$) there must be a reduction in the turbulent nonlinearity \citep{Boldyrev2006,Matthaeus2008_rapid_alignment,Chernoglazov2021_current_structures_in_SRRMHD,Beattie2025_alignment_PRL}, which in our case can possibly be attributed to the strong alignment between velocity $\u$ and vorticity $\bm{\omega}$ (the kinetic helicity), as seen in \citet{Beattie2025_so_long_k41} (in their Appendix, Figure~21) and previously in \citep{Kapyla2018_helicity_SNe_turb}. The presence of an inverse cascade can also cause a shallower spectrum, as shown in detail by \citet{Beattie2025_so_long_k41} which is shown to be possible in helical, hydrodynamical turbulence \citep{Plunian2020_inverse_cascade}, potentially the kind of turbulence that is generated on the expanding SNRs. 

    In the third row we plot the kinetic energy spectra compensated by $k^{-4/3}$, found previously in compressible (magnetized) turbulence \citep{Grete2021_as_a_matter_of_tension}. For $k \lesssim \ell_0^{-1}$, we see close alignment between the $k^{-4/3}$ model and $d\rho u^2(k)/dk$. For SB, we see this behavior at small and large scales, but with a transition between at $k \approx \ell_0^{-1}$. This transition is most likely due to the gaseous profile of the disk imprinting itself across the spectrum. As discussed in the Appendix of  \citet{Beattie2025_so_long_k41}, this increase in kinetic energy density at high $k$, compared to $du^2(k)/dk$, is likely due to high-density clumps on small scales. The velocity power spectrum is more universal across scales and between models, suggesting the velocity modes are correlated through the turbulence in the winds and disk, which we will explore in more detail in \autoref{ssec:universal_cascades}.

    \subsection{What sets the large-scale phase structure in the ISM?} \label{ssec:cs_discussion}
    In the last row of \autoref{fig:powerspec} we show that the sound speed, $c_s \propto \sqrt{T}$, largely scales with the $\u_c$ mode spectrum, demonstrating that the large-scale phase structure of the ISM in the simulations is being controlled directly by the compressible modes, on all $k$ (i.e., that the compressible turbulence is directly responsible for setting the temperature fluctuations in the medium). \citet{Ho2024_multiphase_ISM_turbulence} and \citet{Hu2025_multiphase_ISM_turbulence} have recently found similar results, but \citet{Ho2024_multiphase_ISM_turbulence} relates the effect to the total velocity dispersion,, and \citet{Hu2025_multiphase_ISM_turbulence} argues that the coupling to the ISM is via a Reynolds-averaged diffusion-like process (i.e., $\kappa_T\bnab\cdot[\langle\rho\rangle\bnab\langle T\rangle]$), where $\kappa_T$ is an effective turbulent diffusion transport coefficient that acts to smooth the mean temperature gradients on dynamical timescales $t_{\rm turb}$). We show below that we can completely explain the $c_s$ power spectrum (directly related to the temperature fluctuations), via the $\u_c$ mode fluctuations. 

    Let us consider the general evolution equation for $c_s$ by combining the internal energy and continuity equation,
    \begin{align} \label{eq:phase_structure}
        \frac{\partial c_s^2}{\partial t} + \u\cdot\bnab c_s^2 = \overbrace{-(\gamma - 1) c_s^2 \bnab \cdot \u}^{\substack{\text{adiabatic} \\ \text{contribution}}} \underbrace{- \gamma(\gamma - 1)\frac{\Lambda}{\rho}}_{\substack{\text{cooling function} \\ \text{contribution}}}.
    \end{align}
    The first term on the left-hand-side is the adiabatic contribution, that couples the $\u_c$ modes directly to $c_s$. Intuitively, this term should operate on the timescales of the $\u_c$ modes themselves, which in our model, will be roughly the turbulent eddy turnover time, $\lesssim t_{\rm turb}$ on each scale. The second term is the non-adiabatic contribution from $\Lambda$, which operators on timescales $t_{\rm cool}$. For $c_s = \sqrt{\gamma P/\rho}$, where small pressure and density fluctuations, $\delta \rho$ and $\delta P$, around a background field, $\rho_0$ and $P_0$, respectively, are evolved on $t_{\rm turb}$, such that $t_{\rm turb} \ll t_{\rm cool}$ in the adiabatic limit. Then 
    \begin{align}
        \frac{\delta c_s}{c_{s0}} \sim \frac{1}{2}\left( \frac{\delta P}{P_0} + \frac{\delta \rho}{\rho_0} \right), \quad \text{and}\quad \frac{\delta P }{P_0} \sim \gamma\frac{\delta \rho}{\rho_0},
    \end{align}
    and hence, 
    \begin{align}
        \delta c_s \sim \frac{\gamma - 1}{2}c_{s0}\frac{\delta \rho}{\rho_0}.
    \end{align}
    Using the linearized continuity equation, $-i\omega \Tilde{\delta \rho} + i \rho_0 \k \cdot \Tilde{\u}_c = 0$ where $\omega \sim c_s k $ (i.e., regular acoustic waves) we can rewrite $\delta c_s$ in terms solely of $u_c$,
    \begin{align} \label{eq:cs_uc}
        \Tilde{\delta c}_s \sim \frac{\gamma - 1}{2}c_{s0}\frac{\k\cdot\Tilde{\u}_c}{\omega} \sim \frac{\gamma - 1}{2} \Tilde{u}_c,
    \end{align}
    which means that, in the adiabatic limit where $t_{\rm turb} \ll t_{\rm cool}$, $dc_s^2(k)/dk \sim du_c^2(k)/dk$, as we show in the bottom row\footnote{In \autoref{app:divergence_spectrum} we also show the relation between $du_c^2(k)/dk$ and $d(\bnab\cdot\u)^2(k)/dk$, i.e., $du_c^2(k)/dk \sim d(\bnab\cdot\u)^2(k)/(k^2dk) \sim dc_s^2(k)/dk \propto k^{-2}$.} of \autoref{fig:powerspec}. It should now make sense that $dc_s^2(k)/dk$ for the MW model follows almost exactly the $\u_c$ spectrum, given that the disk has a strong adiabatic component (\autoref{fig:phase}). Any slight deviations are likely due to the existence of non-adiabatic thermodynamic processes. For the wind-dominated SB model, at the disk scale and onward ($\ell_0$ annotated in black dashed line), the spectrum largely follows $k^{-2}$, since the SB disk is also largely adiabatic. For $k \lesssim \ell^{-1}_0$ the spectrum deviates, because the winds are not principally adiabatic and have a mix of thermodynamic processes that we do not account for in our simple adiabatic model. This shows the the important role of compressible turbulent modes in controlling the temperature and phases of the ISMs in our models.

\begin{figure}[htbp]
    \centering
    \includegraphics[width=0.9\linewidth]{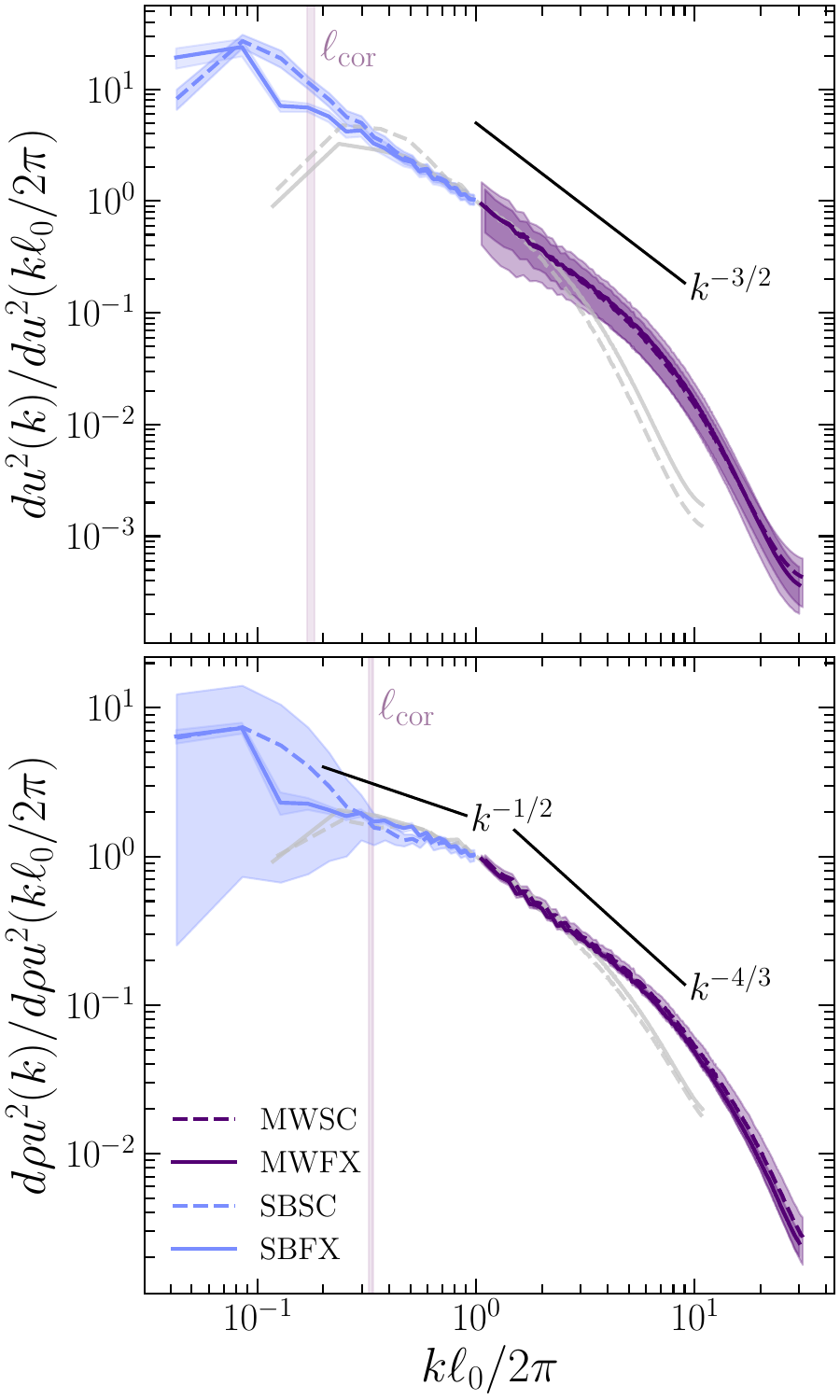}
    \caption[]{
    One-dimensional, isotropic velocity (top) and kinetic energy (bottom) power spectrum, normalized into global gaseous scale heights, $k\ell_0/2\pi = 1$, and power on the scale height (e.g., $du^2(k)/du^2(k\ell_0/2\pi) = 1$) for each galaxy model (MW; dark purple, SB; light purple). We show different SNe seeding schemes (SC; dashed linestyle, FX; solid linestyle). The data that is greyed is either MW, where $k < \ell_0^{-1}$, or SB $k > \ell_0^{-1}$. By normalizing the spectrum into the frame of the scale height we find there is a smooth transition between both the velocity and kinetic energy for the turbulence in the winds and the turbulence in the disk. The velocity spectrum exhibits a universal power law, $du^2(k)/dk \propto k^{-3/2}$, regardless of galaxy model or seeding prescription, whereas the kinetic energy spectrum retains some structure of the galactic disk, varying from $k^{-1/2}$ on $k < \ell_0^{-1}$ to $k^{-4/3}$ on $k > \ell_0^{-1}$. We indicate the global outer scale  $\ell_{\rm{cor}}$ of the $du^2(k)/dk$ and $d\rho u^2(k)/dk$ spectra (\autoref{eq:lcorr}) in the purple shaded line, showing (particularly for the $du^2(k)/dk$ spectrum) that even though we inject momentum from SNe on the smallest scales in the simulations ($\ell_{\rm inj} = 2\rm{dx}$), they drive turbulence to scales well beyond $\ell_0$, with $\ell_{\rm{cor}} \approx 6 \ell_0$.
    }
    \label{fig:K0powerspec}
\end{figure}

    \subsection{Cascading from the winds to the disk}\label{ssec:universal_cascades}
    Based on \autoref{fig:powerspec}, it is clear that many of the spectra seem to follow similar power law scalings across the different seeding and galactic models (which differ in mass, potential depth and injection energy). To explore this universality further, we transform the velocity and kinetic energy spectra such that $k\ell_0/2\pi = 1$ for all models, and $du^2(k)/du^2(k\ell_0/2\pi) = 1$, such that all models are placed in a global gaseous scale height reference frame. We plot the transformed spectra in \autoref{fig:K0powerspec}, showing the SB simulations in light purple (mostly situated in $k\lesssim \ell^{-1}_0$ modes, corresponding to the wind scales), and MW simulations in dark purple (mostly on the $k \gtrsim \ell^{-1}_0$ scales), and the two different seeding schemes with the different linestyles. We change the color of the spectra for the MW models on $k < \ell_0^{-1}$ using grey lines to indicate the spectra beyond those modes, and similarly for the SB spectra but for $k > \ell_0^{-1}$. This allows us to directly see how continuous the spectra are across $k\ell_0/2\pi$, whilst highlighting the structure of the turbulence in the different simulations below and above $k\ell_0/2\pi$.

    We find that all spectra collapse across galactic models and seeding schemes into just a single velocity (top) and kinetic (bottom) spectrum, that are smooth across $\ell^{-1}_0$. The $du^2(k)/dk \propto k^{-3/2}$ discussed in \autoref{ssec:power_spectrum} and previously measured in \citet{Padoan2016_supernova_driving} and \citet{Beattie2025_so_long_k41} extends across both simulations, connecting the winds to the turbulent disk in a continuous manner. For the $d\rho u^2(k)/dk$ spectrum we find a broken power law, that seems to break at roughly $k\ell_0/2\pi \approx 1$, with $d\rho u^2(k)/dk \propto k^{-1/2}$ on $k\ell_0/2\pi \lesssim 1$ and $d\rho u^2(k)/dk \propto k^{-4/3}$ on $k\ell_0/2\pi \gtrsim 1$. This is likely due to the gaseous disk imprinting itself on the power spectrum, due to the mass density contribution in the spectrum. \citet{Federrath2013_universality} previously found that the $d\rho u^2(k)/dk$ was sensitive to the underlying driving field and $\M$, whereas the $du^2(k)/dk$ was not.
    
    We calculate the outer-scale of the turbulence, $\ell_{\rm cor}$,
    \begin{align} \label{eq:lcorr}
        \frac{\ell_{\mathrm{cor}}}{\ell_0} = \frac{1}{\left\langle f^2 \right\rangle}\displaystyle\int_0^\infty dk\; (k\ell_0/2\pi)^{-1} \frac{df^2(k)}{dk},
    \end{align}
    in the global $\ell_0^{-1}$ frame using definitions from \citet{Beattie2025_so_long_k41}, and find $\ell_{\rm cor} \approx 6 \ell_0$ for $du^2(k)/dk$, showing that the largest scale velocity correlations extend well beyond the gaseous scale-height, and $\ell_{\rm cor} \approx3 \ell_0$ for $d\rho u^2(k)/dk$, being truncated at small scales due to the imprint of the galactic disk on $d\rho u^2(k)/dk$. Both the spectra (but particularly the velocity spectrum), show that SNe can drive turbulence to large scales, larger than the conventional wisdom, $\ell_0 \approx \ell_{\rm cor}$ \citep{Haverkorn2008_turbulent_outer_scale,Liu2021_cor_scales,Beattie2022_ion_alfven_fluctuations}. 

\section{Summary and Conclusions}\label{sec:conclusions}
    In this study we analyze the turbulent properties for a variety of $1\,\rm{kpc}^3$ stratified, supernovae (SNe) driven turbulence simulations (two-dimensional slice visualizations shown in \autoref{fig:slice}) with cooling and heating sourced by a time-dependent chemical network, as detailed in \citet{Beattie2025_so_long_k41}. These simulations are relevant for understanding the large-scale properties of the galactic turbulence, probing the volume-filling WIM/WNM ISM phases and the hot galactic winds. Our models are parameterized by their galactic potential, mass, SNe-driving rate, and SNe seed clustering functions, which are previously defined in \citet{Martizzi2016}. These parameterizations allow us to create idealized galactic disk cut-outs, and compare the properties of the turbulence between them. We summarize our main results as follows:
 
    \begin{itemize}
        \item We simulate a small volume of a multiphase Milky Way (MW) analogue and a starburst-like (SB) galaxy (each parameterized by galactic potential, \autoref{eq:phi}, and SNe-driving rate, $\gamma_{\rm{SNe}}$) using two different SNe seeding mechanisms: uniformly at fixed scale height; FX, and mass density-weighted seeding; SC, resulting in four unique models, MWFX, MWSC, SBFX and SBSC. We integrate all models into a statistically stationary state, where the energy injected by the SNe is exactly balanced by the thermalization and (numerical) viscous truncation of the turbulent plasma.
        
        \item We find the volume-averaged turbulent Mach number is $\M \approx 1.7 - 2.0$ (\autoref{fig:mach}), across all simulations, in agreement with observational $\M$ measurements in the WIM of the present-day Milky Way \citep{Gaesnsler_2011_trans_ISM} and low-end starburst galaxies \citep{Westmoquette2009_velocity_dispersion_M82}. Therefore, the volume integral $\M$ is largely insensitive to the $\gamma_{\rm{SNe}}$, mass, galactic potential and seeding mechanism in our simulations.
         
        \item In \autoref{sec:ISM_diversity} we show the notably different dynamical, thermal and density structure in the ISMs between models. The SB models are wind-dominated, characterized by a thin gaseous scale height ($\ell_0\approx40\,{\rm{pc}}$; which we model with an isothermal, pressure-balanced atmosphere model; \autoref{eq:pressure_equilibrium}), a prominent hot galactic wind with large thermal fluctuations, and a short turbulent turnover time on $\ell_0$ of order $t_0\approx0.3\,\rm{Myr}$. The MW model is, on the other hand, disk-dominated with a thick disk ($\ell_0\approx 120$pc), a very tenuous wind with lower thermal and velocity fluctuations, and a longer turbulent turnover time on $\ell_0$ of order $t_0\approx4\,\rm{Myr}$. The FX seeding models lead to a more dynamic, almost adiabatic wind, with larger amplitude thermal and turbulent velocity fluctuations compared to the SC models. 
        
        \item  For each of the models we calculate the power spectrum of the total velocity, kinetic energy, Helmholtz decomposed velocity modes, and sound speed and plot them \autoref{fig:powerspec}. In the disk-dominated MW simulations we show the entire phase structure of the predominantly adiabatic disk can be explained by the $\u_c$ mode turbulence ($dc_s^2(k)/dk \propto du_c^2(k)/dk \propto k^{-2}$), which we can derive by relating $c_s$ and $u_c$ in the weak-cooling, adiabatic limit, assuming that the compressible modes are sourced from turbulence within sound waves, detailed in Equations~\ref{eq:phase_structure}-\ref{eq:cs_uc}. Consistent with \citet{Beattie2025_so_long_k41}, we find a velocity spectrum that is dominated by incompressible modes and close to $du^2(k)/dk \propto k^{-3/2}$, suggesting a weakening of the turbulent nonlinearity which is robust to the different simulation parameters.

        \item By transforming both the velocity $u$ and kinetic energy $\rho u^2
        $ spectrum into units of $\ell_0$ and power on $\ell_0$, in \autoref{fig:K0powerspec} we show that the wind-dominated (starburst; SB) and disk-dominated (Milky Way; MW) galaxy models have power spectra that smoothly transition into one another, suggesting the structure of the turbulence is robust to changes in the galactic potential, SNe-driving rate, and seeding prescription. Furthermore, a universal $du^2(k)/dk \propto k^{-3/2}$ spectrum extends between the disks and the winds seamlessly, suggesting a wind-disk turbulence connection, with the velocity cascade moving through the different ISM phases without hindrance. Furthermore, in this frame, we show that the outer scale of the velocity is $\approx 6\ell_0$, challenging the conventional wisdom that the outer scale is truncated at the gaseous scale height of the disk. The kinetic energy spectrum traces large-scale structure from the galactic disk, with a break scale at $\ell_0^{-1}$, with $k^{-1/2}$ on low $k$ and \textit{approximately}  \citet{Grete2021_as_a_matter_of_tension}'s $k^{-4/3}$ at high $k$. Higher resolution simulations will be required for definitively determining the spectral indexes, but we stress there is no evidence for a Kolmogorov-type $k^{-5/3}$ scaling in either the velocity of kinetic energy spectra.
    \end{itemize}

\section*{\textbf{Acknowledgments}}
    We thank the anonymous reviewer for helping increase the clarity and presentation of the study. We further thank Norm Murray, Peng Oh, Jim Stone, Ralf Klessen, Christoph Federrath, and the CITA plasma-astro group for the many enlightening discussions. I.~C. and J.~R.~B. acknowledge compute allocations rrg-ripperda and rrg-essick from the Digital Research Alliance of Canada, which were utilized for both running and analyzing the simulations. I.~C. acknowledges support from the Barry M. Goldwater Scholarship and acknowledges funding from the UC Santa Cruz Undergraduate Research in Science \& Technology (URST) award. J.~R.~B. acknowledges funding from the Natural Sciences and Engineering Research Council of Canada (NSERC) (funding reference number 568580) and acknowledges the support from NSF Award 2206756, as well as high-performance computing resources provided by the Leibniz Rechenzentrum and the Gauss Center for Supercomputing grants~pn76gi,~pr73fi and pn76ga. 
    This research was made possible thanks to funding from the Lamat Institute and UC Santa Cruz through the Heising-Simons Foundation and NSF grants: AST-1852393, AST-1911206, AST-2150255 and AST-2206243.

    \facilities{All simulations were run and processed on the two Digital Research Alliance of Canada supercomputers, Niagara and Trillium.}

    \software{We use \textsc{ramses} \citep{Teyssier2002_ramses} for all of the simulations. Data analysis and visualization software used in this study: \textsc{C++} \citep{Stroustrup2013}, \textsc{numpy} \citep{Oliphant2006,numpy2020}, \textsc{numba}, \citep{numba:2015}, \textsc{matplotlib} \citep{Hunter2007}, \textsc{cython} \citep{Behnel2011}, \textsc{visit} \citep{Childs2012}, \textsc{scipy} \citep{Virtanen2020}, \textsc{scikit-image} \citep{vanderWalts2014}, \textsc{cmasher} \citep{Velden2020_cmasher}, \textsc{yt} \citep{yt}, \textsc{pandas} \citep{pandas}, \textsc{joblib}\citep{joblib}, \textsc{pyfftw}\citep{2021ascl.soft09009G}}
\appendix

\begin{figure*}[htbp]
    \centering
    \includegraphics[width=\linewidth]{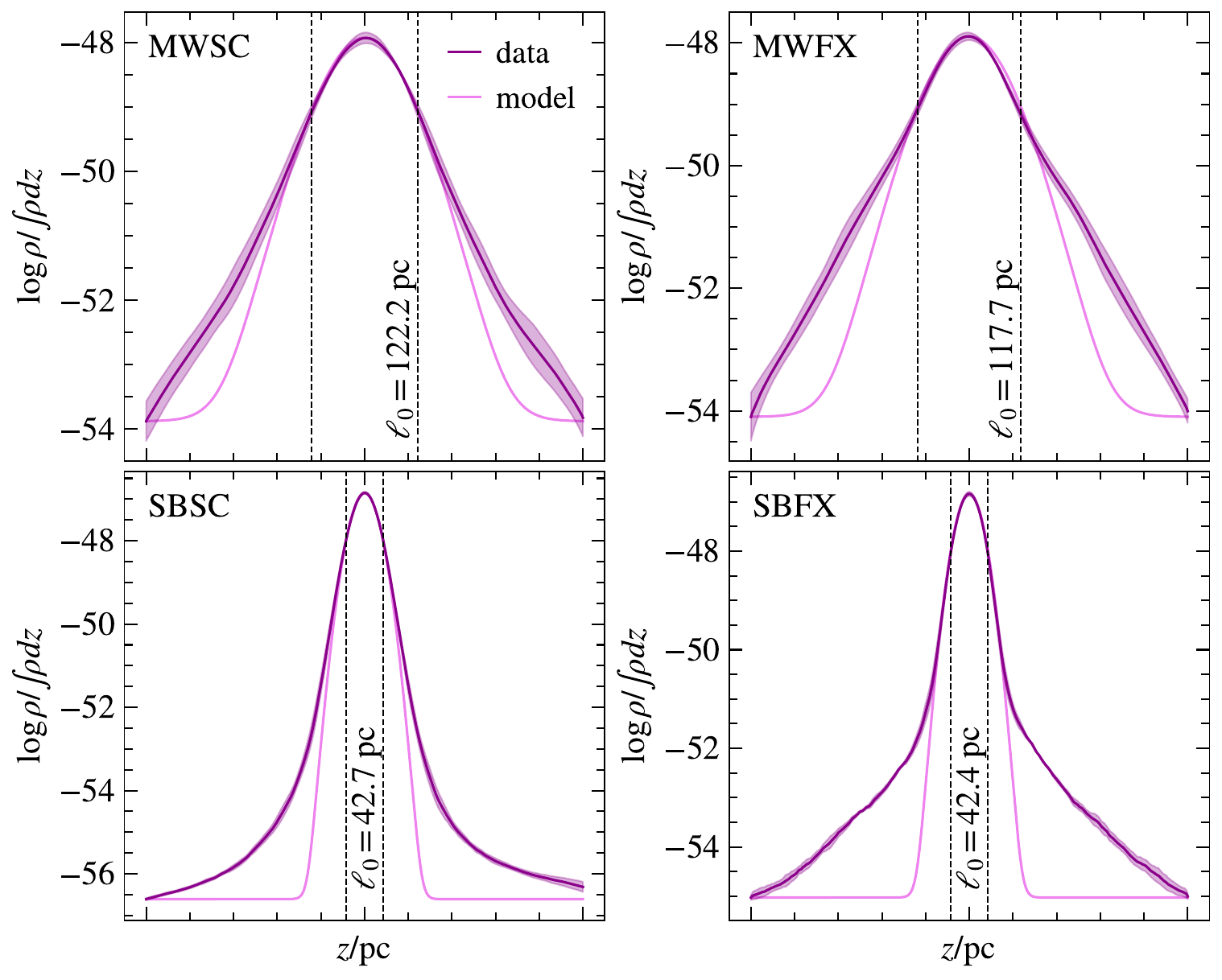}
    \caption[]{Vertical disk profile fits (light purple) for each model (top left annotation; dark purple; also shown in \autoref{fig:4panel_density}) using the isothermal pressure-equilibrium profile model (\autoref{eq:isothermal_profile}), with free parameter $\sigma_z^2$. We use this disk-component model to calculate gaseous scale height $\ell_0$ (\autoref{eq:scaleheight}) detailed in \autoref{ssec:scaleheight}. To fit, we use maximum likelihood fitting, constructing the posterior by utilizing \textsc{emcee} \citep{emcee}, method detailed in \autoref{app:fitting}.}
    \label{fig:4panel_density_app}
\end{figure*}

\section{Error propagation and fitting the galactic profile models} \label{app:fitting}
    We model the vertical density profile of the disk using \autoref{eq:isothermal_profile} (detailed in \autoref{ssec:scaleheight}) and fit the free parameter $\sigma_z^2$ (\autoref{eq:sigma_z}). We employ a MCMC maximum-likelihood method using the emcee \citep{emcee} ensemble sampler to construct the posterior. The likelihood was constructed from the log-density profiles $\log\rho(z)$, weighted by the variance at each height $z$, and an additional Gaussian window function to extract only the disk component. We pick uniform priors: $10^4\,\rm{cms}^{-1}<\sigma_z<8\times10^6\,\rm{cms}^{-1}$, based on physical constraints for the turbulent and thermally supported disk. For each model we ran $50$ chains for $5,\!000$ steps, discarding the first $20\%$ of the chains as burn-in stage. We report the median and standard deviation of $\sigma_z$ as the best-fit value and uncertainty directly from the posterior. From $\sigma_z$ and \autoref{eq:scaleheight}, we derive the gaseous scale height $\ell_0$, and propagate the uncertainty by applying standard Gaussian error propagation, where
    \begin{align}
        \sigma_{\ell_0} = \left|\frac{\partial\ell_0}{\partial\sigma_z}\right|\sigma_{\sigma_z}, \quad
        \zeta= \frac{\sigma_z^2}{a_1} + z_0,\quad \ell_0=\sqrt{\zeta^2 - z_0^2}, \quad\text{and}\quad
        \frac{\partial\ell_0}{\partial\sigma_z} =\frac{\zeta}{\sqrt{\zeta^2-z_0^2}}\frac{2\sigma_z}{a_1},
    \end{align}
    and $a_1$ is the acceleration of the stellar disk. Putting this together,
    \begin{align}
        \sigma_{\ell_0}=\left(\frac{2\sigma_z u}{a_1\sqrt{u^2-z_0^2}}\right)\sigma_{\sigma_z}.
    \end{align}
    which we report for each model in Table~\ref{tb:results}.

\section{Fitted Power Law Indexes For Spectra}\label{app:slopes}
\begin{deluxetable}{cccccc}\label{tb:slopes}
\tablecaption{Fitted power laws of the one-dimensional, isotropic power spectra quantities normalized by total power, as a function
of wavenumber, $k$, normalized by the system scale wavenumber, $2\pi/L$, in \autoref{fig:powerspec} for each model. }
\tablehead{
\colhead{Model} &
\colhead{$du^2(k)/dk$} &
\colhead{$du_s^2(k)/dk$} &
\colhead{$du_c^2(k)/dk$} &
\colhead{$d\rho u^2(k)/dk$} & 
\colhead{$dc_s^2(k)/dk$} \\
\colhead{(1)} & \colhead{(2)} & \colhead{(3)} & \colhead{(4)} & \colhead{(5)} & \colhead{(6)}
}

\startdata
MWSC & $-1.48\pm0.12$ & $-1.30\pm0.11$ & $-1.94\pm 0.17$ & $-1.16\pm0.05$ & $-1.83\pm0.10$ \\
MWFX & $-1.43\pm0.14 $ & $-1.26\pm0.12 $ & $-1.95\pm 0.21$ & $-1.19\pm0.04$ & $-1.73\pm0.12$\\
SBSC &$-1.24\pm0.05$ & $-1.07\pm0.05$ & $-1.73\pm0.12$ & $-0.39\pm0.14$ & $-1.20\pm0.07$ \\
SBFX &$-1.14\pm0.04$ & $-1.01\pm0.03$ & $-1.59\pm0.10$ & $-0.61\pm0.03$ & $-1.19\pm0.03$ \\
\hline\\[-1.25em]
\enddata
\tablecomments{\textbf{Column (1)}: Numerical model label, where MW means Milky Way analogue, SB means starburst analogue, SC refers to the self-consistent SNe seeding prescription, and FX refers to the fixed seeding prescription (see \autoref{sssec:sc} for more details about the exact seeding prescription). All indexes are derived from fits across the range of modes $kL/2\pi \in [8.5,25]$. \textbf{Column (2)}: The velocity power spectrum. \textbf{Column (3)}: The incompressible velocity mode power spectrum. \textbf{Column (4)}: The compressible velocity mode power spectrum. \textbf{Column (5)}: The kinetic energy power spectrum. \textbf{Column (6)}: The sound speed power spectrum. }
\end{deluxetable}

    In Table~\ref{tb:slopes} we show the power law indexes derived by fitting to the scaling ranges of each of the spectra discussed in \autoref{sec:power_spectrum} and plotted in \autoref{fig:powerspec}. For the fits, we use least-squares in $\log-\log$, propagating the $1\sigma$ from the temporal averaging for each of the spectra. As shown in \citet{Beattie2025_so_long_k41}, it is not obvious that there is a true inertial (constant $u^3/\ell$) range of modes, so we therefore just pick the range $kL/2\pi\in [8.5, 25]$ by eye, and perform the fit over that range. Because these simulations are not scale-free, i.e., the box is stratified with scale heights in all quantities (as we show in \autoref{fig:4panel_density} and \autoref{fig:4panel_density_app}), we are unable to directly compare the indexes between the galaxy models because $kL/2\pi \in [8.5, 25]$ is probing different physical scales in each of the simulations (see \autoref{fig:K0powerspec} for how to properly combine and compare the spectra across galaxies). However, for the same galactic models, with different seeds, we see broad agreement within $1\sigma$, as expected qualitatively from \autoref{fig:powerspec}.

\section{Divergence and Compressible Mode Power Spectrum}\label{app:divergence_spectrum}

\begin{figure*}[htbp]
    \centering
    \includegraphics[width=\linewidth]{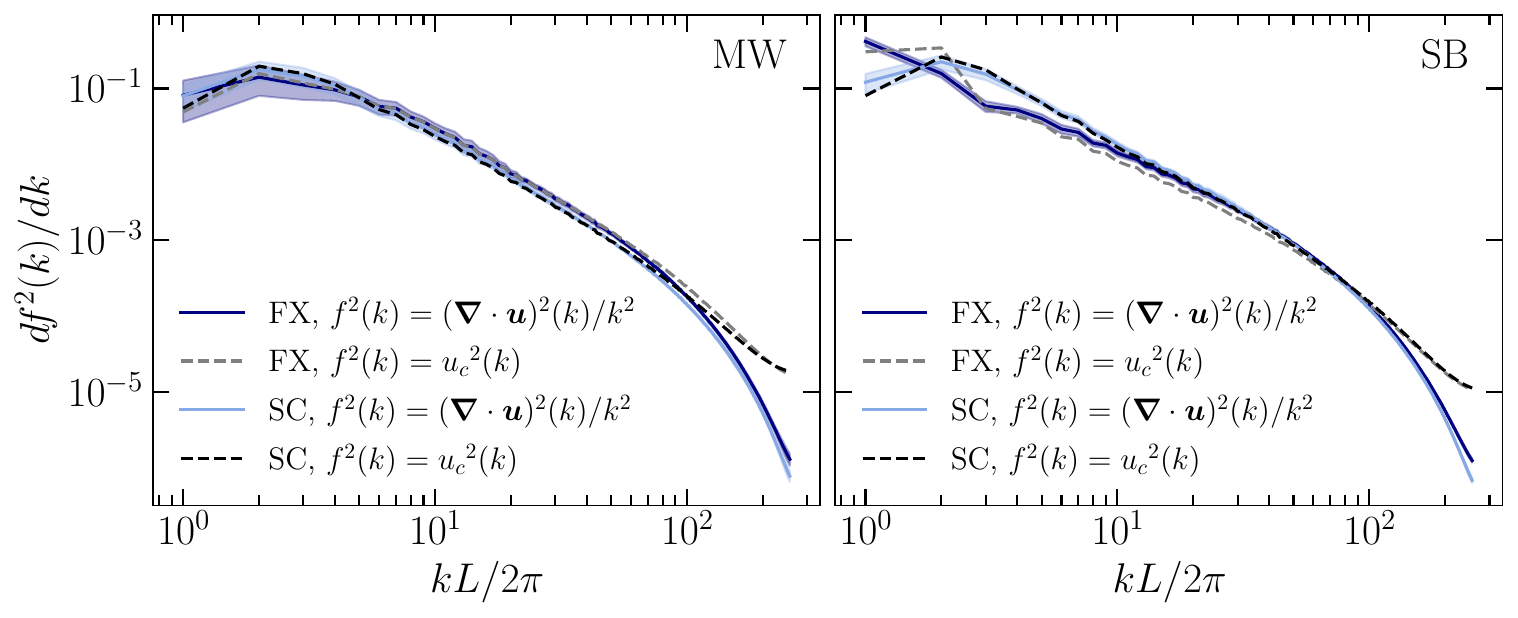}
    \caption[]{One-dimensional, isotropic velocity divergence, $\bm{\nabla}\cdot\bm{u}$, power spectrum (blues), and compressible mode, $\u_c$, power spectrum (dashed lines) normalized by the integrated power, for each seeding mechanism, for MW (left) and SB (right) galaxy models. We presented in \autoref{ssec:cs_discussion} and \autoref{fig:powerspec} (fourth row) that the sound speed scales with the compressible mode spectrum, explaining the temperature fluctuations in the medium. Further, as expected, we show that the divergence spectrum and the compressible mode spectrum are simply related by a $k^2$ scaling to get the divergence into the correct dimensions.}
    \label{fig:powerspec_appendix}
\end{figure*}

    In \autoref{sec:power_spectrum} we show the compressible mode $\u_c$ spectrum was directly related to the sound speed spectrum, which was controlling the ISM phase structure in our models with a thick disk (Milky Way analogue models). Here we formalize the connection with $\u_c$ and the fluid divergence. The divergence $\bnab\cdot\u$ and $\u_c$ are intrinsically related because only $\u_c$ satisfies $\bnab\cdot\u \neq 0$. In \autoref{fig:powerspec_appendix} we plot the power spectrum of both quantities (left MW models, right SB models), with the expectation that $du_c^2(k)/dk \propto d(\bnab\cdot\u)^2(k)/(k^2dk)$ based on dimensionality. Indeed, we find that $du_c^2(k)/dk \propto d(\bnab\cdot\u)^2(k)/(k^2dk)$ across all resolved modes in the simulations. As we show in \autoref{fig:powerspec}, $du_c^2(k)/dk \propto k^{-2}$, which we associate with \citet{Burgers1948} turbulence (as in \citealt{Beattie2025_nature_astronomy} on large scales). 

\begin{figure*}[htbp]
    \centering
    \includegraphics[width=\linewidth]{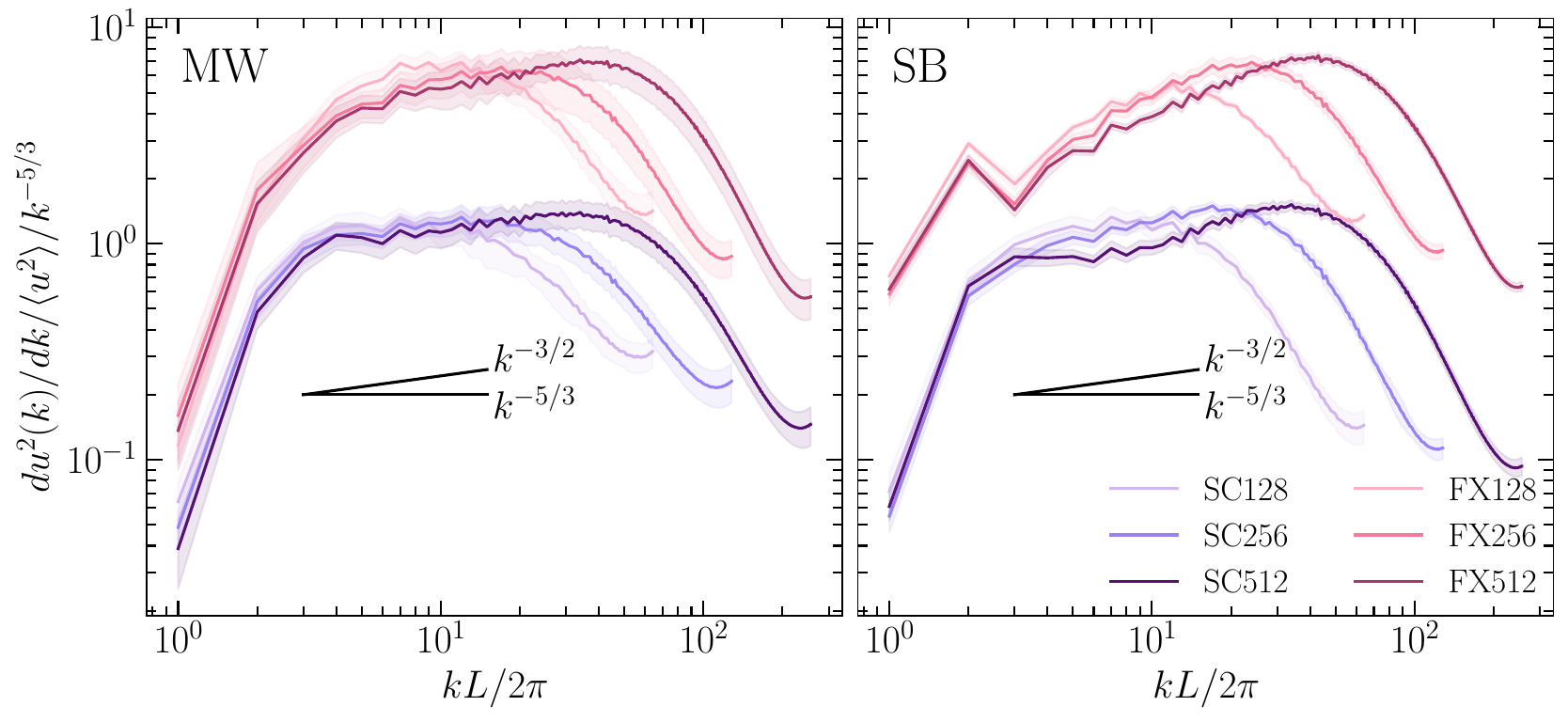}
    \caption[]{One-dimensional, isotropic velocity power spectra, $du^2(k)/dk$, normalized by the integrated power, as a function of wavenumber, $k$, normalized by the system scale wavenumber, $2\pi/L$. We compensate the spectra by $k^{-5/3}$ to compare to the \citealt{Kolmogorov1941} power spectrum. MW runs are shown on the left panel and SB on the right, with each color indicating a different grid resolution of the FX (pinks) and SC (purples) seeding schemes.}
    \label{fig:convergence}
\end{figure*}

\section{Convergence test}\label{app:convergence}

    For the purpose of testing the convergence of our spectral results, we calculate the velocity power spectra at each resolution ($N_{\rm grid}^3 = 128^3,\; 256^3,\; \text{and}\; 512^3$) for all models. We show this in \autoref{fig:convergence}, with MW models on the left, and SB models on the right, and SC and FX runs shown in purples and pinks, respectively. We normalize by the integral power and compensate by $k^{-5/3}$, as in \autoref{sec:power_spectrum}. It is visible that as we increase the resolution, we resolve an extended range of self-similar modes in the spectrum, increasing the power at higher $k$ as the numerical dissipation scale moves to smaller scales. For implicit large-eddy simulations, this is exactly the behavior we expect because the viscous truncation of the cascade is purely a function of the grid \citep{Grete2023_as_a_matter_of_dynamical_range,Shivakumar2025_numerical_dissipation,Grehan2025_numerical_reconnection}. We find $128^3$ is insufficient to resolve any of the self-similar range of modes, but as we go to higher resolutions, we find that the self-similar range of modes emerges, and at $512^3$ we have almost an order of magnitude of modes in this self-similar range. Furthermore, the spectra begin to roughly follow $\sim k^{-3/2}$. The key point is that the spectra begin to converge to a similar shape as we go to higher resolution, which means that the measurements we make in this study are capturing some of the asymptotic dynamics of the SNe-driven turbulence.

\newpage
\bibliography{Mar24,james,izzy}{}

\begin{thebibliography}{}
\expandafter\ifx\csname natexlab\endcsname\relax\def\natexlab#1{#1}\fi
\providecommand{\url}[1]{\href{#1}{#1}}
\providecommand{\dodoi}[1]{doi:~\href{http://doi.org/#1}{\nolinkurl{#1}}}
\providecommand{\doeprint}[1]{\href{http://ascl.net/#1}{\nolinkurl{http://ascl.net/#1}}}
\providecommand{\doarXiv}[1]{\href{https://arxiv.org/abs/#1}{\nolinkurl{https://arxiv.org/abs/#1}}}

\bibitem[{{Adebahr} {et~al.}(2013){Adebahr}, {Krause}, {Klein}, {We{\.z}gowiec}, {Bomans}, \& {Dettmar}}]{adebahr_m82_scaleheight}
{Adebahr}, B., {Krause}, M., {Klein}, U., {et~al.} 2013, \aap, 555, A23, \dodoi{10.1051/0004-6361/201220226}

\bibitem[{{Balsara} {et~al.}(2004){Balsara}, {Kim}, {Mac Low}, \& {Mathews}}]{Balsara2004_SNe_turbulence_and_dynamo}
{Balsara}, D.~S., {Kim}, J., {Mac Low}, M.-M., \& {Mathews}, G.~J. 2004, \apj, 617, 339, \dodoi{10.1086/425297}

\bibitem[{{Beattie} \& {Bhattacharjee}(2025)}]{Beattie2025_alignment_PRL}
{Beattie}, J.~R., \& {Bhattacharjee}, A. 2025, arXiv e-prints, arXiv:2504.15538, \dodoi{10.48550/arXiv.2504.15538}

\bibitem[{{Beattie} {et~al.}(2025{\natexlab{a}}){Beattie}, {Federrath}, {Klessen}, {Cielo}, \& {Bhattacharjee}}]{Beattie2025_nature_astronomy}
{Beattie}, J.~R., {Federrath}, C., {Klessen}, R.~S., {Cielo}, S., \& {Bhattacharjee}, A. 2025{\natexlab{a}}, Nature Astronomy, \dodoi{10.1038/s41550-025-02551-5}

\bibitem[{{Beattie} {et~al.}(2025{\natexlab{b}}){Beattie}, {Federrath}, {Kriel}, {Hew}, \& {Bhattacharjee}}]{Beattie2025_compressible_dynamo}
{Beattie}, J.~R., {Federrath}, C., {Kriel}, N., {Hew}, J. K.~J., \& {Bhattacharjee}, A. 2025{\natexlab{b}}, \mnras, \dodoi{10.1093/mnras/staf1318}

\bibitem[{{Beattie} {et~al.}(2022{\natexlab{a}}){Beattie}, {Krumholz}, {Federrath}, {Sampson}, \& {Crocker}}]{Beattie2022_ion_alfven_fluctuations}
{Beattie}, J.~R., {Krumholz}, M.~R., {Federrath}, C., {Sampson}, M.~L., \& {Crocker}, R.~M. 2022{\natexlab{a}}, Frontiers in Astronomy and Space Sciences, 9, 900900, \dodoi{10.3389/fspas.2022.900900}

\bibitem[{{Beattie} {et~al.}(2022{\natexlab{b}}){Beattie}, {Mocz}, {Federrath}, \& {Klessen}}]{Beattie2022_spdf}
{Beattie}, J.~R., {Mocz}, P., {Federrath}, C., \& {Klessen}, R.~S. 2022{\natexlab{b}}, The Monthly Notices of The Royal Astronomical Society, 517, 5003, \dodoi{10.1093/mnras/stac3005}

\bibitem[{{Beattie} {et~al.}(2025{\natexlab{c}}){Beattie}, {Noer Kolborg}, {Ramirez-Ruiz}, \& {Federrath}}]{Beattie2025_so_long_k41}
{Beattie}, J.~R., {Noer Kolborg}, A., {Ramirez-Ruiz}, E., \& {Federrath}, C. 2025{\natexlab{c}}, arXiv e-prints, arXiv:2501.09855, \dodoi{10.48550/arXiv.2501.09855}

\bibitem[{{Beck} \& {Wielebinski}(2013)}]{Beck_2013_Bfield_in_gal}
{Beck}, R., \& {Wielebinski}, R. 2013, {Magnetic Fields in Galaxies}, ed. T.~D. {Oswalt} \& G.~{Gilmore}, Vol.~5, 641, \dodoi{10.1007/978-94-007-5612-0\_13}

\bibitem[{Behnel {et~al.}(2011)Behnel, Bradshaw, Citro, Dalcin, Seljebotn, \& Smith}]{Behnel2011}
Behnel, S., Bradshaw, R., Citro, C., {et~al.} 2011, Computing in Science \& Engineering, 13, 31

\bibitem[{Boldyrev(2006)}]{Boldyrev2006}
Boldyrev, S. 2006, Physical Review Letters, 96, 115002, \dodoi{10.1103/PhysRevLett.96.115002}

\bibitem[{Burgers(1948)}]{Burgers1948}
Burgers, J. 1948, Advances in Applied Mechanics, 1, 171, \dodoi{http://dx.doi.org/10.1016/S0065-2156(08)70100-5}

\bibitem[{{Burkhart}(2018)}]{Burkhart2018}
{Burkhart}, B. 2018, The Astrophysical Journal, 863, 118, \dodoi{10.3847/1538-4357/aad002}

\bibitem[{Burkhart {et~al.}(2009)Burkhart, Falceta-Gon{\c{c}}alves, Kowal, \& Lazarian}]{Burkhart2009_bispectrum}
Burkhart, B., Falceta-Gon{\c{c}}alves, D., Kowal, G., \& Lazarian, A. 2009, The Astrophysical Journal, 693, 250, \dodoi{10.1088/0004-637x/693/1/250}

\bibitem[{{Burkhart} \& {Mocz}(2019)}]{Burkhart2019}
{Burkhart}, B., \& {Mocz}, P. 2019, \apj, 879, 129, \dodoi{10.3847/1538-4357/ab25ed}

\bibitem[{{Burkhart} {et~al.}(2020){Burkhart}, {Appel}, {Bialy}, {Cho}, {Christensen}, {Collins}, {Federrath}, {Fielding}, {Finkbeiner}, {Hill}, {Ib{\'a}{\~n}ez-Mej{\'\i}a}, {Krumholz}, {Lazarian}, {Li}, {Mocz}, {Mac Low}, {Naiman}, {Portillo}, {Shane}, {Slepian}, \& {Yuan}}]{Burkhart2020_CATS}
{Burkhart}, B., {Appel}, S.~M., {Bialy}, S., {et~al.} 2020, \apj, 905, 14, \dodoi{10.3847/1538-4357/abc484}

\bibitem[{{Chamandy} \& {Shukurov}(2020)}]{2020Galax...8...56C}
{Chamandy}, L., \& {Shukurov}, A. 2020, Galaxies, 8, 56, \dodoi{10.3390/galaxies8030056}

\bibitem[{{Chernoglazov} {et~al.}(2021){Chernoglazov}, {Ripperda}, \& {Philippov}}]{Chernoglazov2021_current_structures_in_SRRMHD}
{Chernoglazov}, A., {Ripperda}, B., \& {Philippov}, A. 2021, The Astrophysical Journal Letters, 923, L13, \dodoi{10.3847/2041-8213/ac3afa}

\bibitem[{{Chevalier} \& {Clegg}(1985)}]{Chevalier_1985_Nature}
{Chevalier}, R.~A., \& {Clegg}, A.~W. 1985, \nat, 317, 44, \dodoi{10.1038/317044a0}

\bibitem[{Childs {et~al.}(2012)Childs, Brugger, Whitlock, Meredith, Ahern, Pugmire, Biagas, Miller, Harrison, Weber, Krishnan, Fogal, Sanderson, Garth, Bethel, Camp, R\"{u}bel, Durant, Favre, \& Navr\'{a}til}]{Childs2012}
Childs, H., Brugger, E., Whitlock, B., {et~al.} 2012, in {High Performance Visualization--Enabling Extreme-Scale Scientific Insight} (Taylor \& Francis), 357--372

\bibitem[{{Clark} \& {Hensley}(2019)}]{Clark2019}
{Clark}, S.~E., \& {Hensley}, B.~S. 2019, The Astrophysical Journal, 887, 136, \dodoi{10.3847/1538-4357/ab5803}

\bibitem[{{Clark} {et~al.}(2015){Clark}, {Hill}, {Peek}, {Putman}, \& {Babler}}]{Clark2015_cold_structures}
{Clark}, S.~E., {Hill}, J.~C., {Peek}, J.~E.~G., {Putman}, M.~E., \& {Babler}, B.~L. 2015, \prl, 115, 241302, \dodoi{10.1103/PhysRevLett.115.241302}

\bibitem[{Conroy \& Kratter(2012)}]{Conroy_2012_runawayOB}
Conroy, C., \& Kratter, K.~M. 2012, The Astrophysical Journal, 755, 123, \dodoi{10.1088/0004-637X/755/2/123}

\bibitem[{{de Avillez} \& {Breitschwerdt}(2005)}]{deAvillez2005_ISM_phases_SNe}
{de Avillez}, M.~A., \& {Breitschwerdt}, D. 2005, \aap, 436, 585, \dodoi{10.1051/0004-6361:20042146}

\bibitem[{{de Wit} {et~al.}(2004){de Wit}, {Testi}, {Palla}, {Vanzi}, \& {Zinnecker}}]{deWit_2004_OB_clustering}
{de Wit}, W.~J., {Testi}, L., {Palla}, F., {Vanzi}, L., \& {Zinnecker}, H. 2004, \aap, 425, 937, \dodoi{10.1051/0004-6361:20040454}

\bibitem[{{Dhawalikar} {et~al.}(2022){Dhawalikar}, {Federrath}, {Davidovits}, {Teyssier}, {Nagel}, {Remington}, \& {Collins}}]{Dhawalikar2022_shock_driving_parameter}
{Dhawalikar}, S., {Federrath}, C., {Davidovits}, S., {et~al.} 2022, The Monthly Notices of The Royal Astronomical Society, 514, 1782, \dodoi{10.1093/mnras/stac1480}

\bibitem[{{Diehl} {et~al.}(2006){Diehl}, {Halloin}, {Kretschmer}, {Lichti}, {Sch{\"o}nfelder}, {Strong}, {von Kienlin}, {Wang}, {Jean}, {Kn{\"o}dlseder}, {Roques}, {Weidenspointner}, {Schanne}, {Hartmann}, {Winkler}, \& {Wunderer}}]{Diehl2006_SNe_rate_MW}
{Diehl}, R., {Halloin}, H., {Kretschmer}, K., {et~al.} 2006, \nat, 439, 45, \dodoi{10.1038/nature04364}

\bibitem[{Elmegreen \& Scalo(2004)}]{Elmegreen2004}
Elmegreen, B.~G., \& Scalo, J. 2004, Annu. Rev. Astron. Astrophys., 42, 211, \dodoi{10.1146/annurev.astro.41.011802.094859}

\bibitem[{{Elmegreen} {et~al.}(2025){Elmegreen}, {Calzetti}, {Adamo}, {Sandstrom}, {Dale}, {Bajaj}, {Boyer}, {Duarte-Cabral}, {Chown}, {Correnti}, {Dalcanton}, {Draine}, {Gaches}, {Gallagher}, {Grasha}, {Gregg}, {Hunt}, {Johnson}, {Kennicutt}, {Klessen}, {Leroy}, {Linden}, {McLeod}, {Messa}, {{\"O}stlin}, {Padave}, {Roman-Duval}, {Smith}, {Walter}, \& {Weinbeck}}]{elmegreen_2025_scaleheight_SB}
{Elmegreen}, B.~G., {Calzetti}, D., {Adamo}, A., {et~al.} 2025, \apj, 986, 13, \dodoi{10.3847/1538-4357/adcee6}

\bibitem[{{Ewart} {et~al.}(2025){Ewart}, {Reichherzer}, {Ren}, {Majeski}, {Mori}, {Nastac}, {Bott}, {Kunz}, \& {Schekochihin}}]{Ewart2025_CR_multiphase_medium}
{Ewart}, R.~J., {Reichherzer}, P., {Ren}, S., {et~al.} 2025, arXiv e-prints, arXiv:2507.19044, \dodoi{10.48550/arXiv.2507.19044}

\bibitem[{Federrath(2013)}]{Federrath2013_universality}
Federrath, C. 2013, The Monthly Notices of The Royal Astronomical Society, 436, 1245, \dodoi{10.1093/mnras/stt1644}

\bibitem[{{Federrath}(2015)}]{Federrath2015_inefficient_SFR}
{Federrath}, C. 2015, The Monthly Notices of The Royal Astronomical Society, 450, 4035, \dodoi{10.1093/mnras/stv941}

\bibitem[{Federrath \& Klessen(2012)}]{Federrath2012}
Federrath, C., \& Klessen, R.~S. 2012, The Astrophysical Journal, 761, \dodoi{10.1088/0004-637X/761/2/156}

\bibitem[{{Federrath} {et~al.}(2008){Federrath}, {Klessen}, \& {Schmidt}}]{Federrath2008}
{Federrath}, C., {Klessen}, R.~S., \& {Schmidt}, W. 2008, The Astrophysical Journal Letters, 688, L79, \dodoi{10.1086/595280}

\bibitem[{{Federrath} {et~al.}(2016){Federrath}, {Rathborne}, {Longmore}, {Kruijssen}, {Bally}, {Contreras}, {Crocker}, {Garay}, {Jackson}, {Testi}, \& {Walsh}}]{Federrath2016_brick}
{Federrath}, C., {Rathborne}, J.~M., {Longmore}, S.~N., {et~al.} 2016, The Astrophysical Journal, 832, 143, \dodoi{10.3847/0004-637X/832/2/143}

\bibitem[{Ferrand {et~al.}(2020)Ferrand, Galtier, Sahraoui, \& Federrath}]{Ferrand_2020}
Ferrand, R., Galtier, S., Sahraoui, F., \& Federrath, C. 2020, The Astrophysical Journal, 904, 160, \dodoi{10.3847/1538-4357/abb76e}

\bibitem[{{Ferri{\`e}re}(2020)}]{Ferriere2020_reynolds_numbers_for_ism}
{Ferri{\`e}re}, K. 2020, Plasma Physics and Controlled Fusion, 62, 014014, \dodoi{10.1088/1361-6587/ab49eb}

\bibitem[{Fielding {et~al.}(2018)Fielding, Quataert, \& Martizzi}]{Fielding_winds}
Fielding, D., Quataert, E., \& Martizzi, D. 2018, Monthly Notices of the Royal Astronomical Society, 481, 3325, \dodoi{10.1093/mnras/sty2466}

\bibitem[{{Fielding} {et~al.}(2023){Fielding}, {Ripperda}, \& {Philippov}}]{Fielding2022_ISM_plasmoids}
{Fielding}, D.~B., {Ripperda}, B., \& {Philippov}, A.~A. 2023, The Astrophysical Journal Letters, 949, L5, \dodoi{10.3847/2041-8213/accf1f}

\bibitem[{{Foreman-Mackey} {et~al.}(2013){Foreman-Mackey}, {Hogg}, {Lang}, \& {Goodman}}]{emcee}
{Foreman-Mackey}, D., {Hogg}, D.~W., {Lang}, D., \& {Goodman}, J. 2013, PASP, 125, 306, \dodoi{10.1086/670067}

\bibitem[{{Gaensler} {et~al.}(2011{\natexlab{a}}){Gaensler}, {Haverkorn}, {Burkhart}, {Newton-McGee}, {Ekers}, {Lazarian}, {McClure-Griffiths}, {Robishaw}, {Dickey}, \& {Green}}]{Gaesnsler_2011_trans_ISM}
{Gaensler}, B.~M., {Haverkorn}, M., {Burkhart}, B., {et~al.} 2011{\natexlab{a}}, \nat, 478, 214, \dodoi{10.1038/nature10446}

\bibitem[{{Gaensler} {et~al.}(2011{\natexlab{b}}){Gaensler}, {Haverkorn}, {Burkhart}, {Newton-McGee}, {Ekers}, {Lazarian}, {McClure-Griffiths}, {Robishaw}, {Dickey}, \& {Green}}]{Gaensler2011}
---. 2011{\natexlab{b}}, \nat, 478, 214, \dodoi{10.1038/nature10446}

\bibitem[{Galbany {et~al.}(2018)Galbany, Anderson, Sánchez, Kuncarayakti, Pedraz, González-Gaitán, Stanishev, Domínguez, Moreno-Raya, Wood-Vasey, Mourão, Ponder, Badenes, Mollá, López-Sánchez, Rosales-Ortega, Vílchez, García-Benito, \& Marino}]{Galbany_2018_SNE_dense}
Galbany, L., Anderson, J.~P., Sánchez, S.~F., {et~al.} 2018, The Astrophysical Journal, 855, 107, \dodoi{10.3847/1538-4357/aaaf20}

\bibitem[{{Gallegos-Garcia} {et~al.}(2020){Gallegos-Garcia}, {Burkhart}, {Rosen}, {Naiman}, \& {Ramirez-Ruiz}}]{Gallegos-Garcia2020}
{Gallegos-Garcia}, M., {Burkhart}, B., {Rosen}, A.~L., {Naiman}, J.~P., \& {Ramirez-Ruiz}, E. 2020, \apjl, 899, L30, \dodoi{10.3847/2041-8213/ababae}

\bibitem[{{Gatto} {et~al.}(2015){Gatto}, {Walch}, {Low}, {Naab}, {Girichidis}, {Glover}, {W{\"u}nsch}, {Klessen}, {Clark}, {Baczynski}, {Peters}, {Ostriker}, {Ib{\'a}{\~n}ez-Mej{\'\i}a}, \& {Haid}}]{Gatto2015_SNe_clustering}
{Gatto}, A., {Walch}, S., {Low}, M. M.~M., {et~al.} 2015, \mnras, 449, 1057, \dodoi{10.1093/mnras/stv324}

\bibitem[{{Gent} {et~al.}(2021){Gent}, {Mac Low}, {K{\"a}pyl{\"a}}, \& {Singh}}]{Gent2021_supernova_turbulence_and_dynamo}
{Gent}, F.~A., {Mac Low}, M.-M., {K{\"a}pyl{\"a}}, M.~J., \& {Singh}, N.~K. 2021, \apjl, 910, L15, \dodoi{10.3847/2041-8213/abed59}

\bibitem[{{Gent} {et~al.}(2023){Gent}, {Mac Low}, {Korpi-Lagg}, \& {Singh}}]{Gent2022_multiphase_dynamo}
{Gent}, F.~A., {Mac Low}, M.-M., {Korpi-Lagg}, M.~J., \& {Singh}, N.~K. 2023, The Astrophysical Journal, 943, 176, \dodoi{10.3847/1538-4357/acac20}

\bibitem[{Gent {et~al.}(2013)Gent, Shukurov, Fletcher, Sarson, \& Mantere}]{Gent_2017}
Gent, F.~A., Shukurov, A., Fletcher, A., Sarson, G.~R., \& Mantere, M.~J. 2013, Monthly Notices of the Royal Astronomical Society, 432, 1396, \dodoi{10.1093/mnras/stt560}

\bibitem[{{Gerrard} {et~al.}(2023){Gerrard}, {Federrath}, {Pingel}, {McClure-Griffiths}, {Marchal}, {Joncas}, {Clark}, {Stanimirovi{\'c}}, {Lee}, {van Loon}, {Dickey}, {D{\'e}nes}, {Ma}, {Dempsey}, \& {Lynn}}]{Gerrard2023_LMC}
{Gerrard}, I.~A., {Federrath}, C., {Pingel}, N.~M., {et~al.} 2023, \mnras, 526, 982, \dodoi{10.1093/mnras/stad2718}

\bibitem[{{Gomersall}(2021)}]{2021ascl.soft09009G}
{Gomersall}, H. 2021, {pyFFTW: Python wrapper around FFTW}, Astrophysics Source Code Library, record ascl:2109.009

\bibitem[{{Grehan} {et~al.}(2025){Grehan}, {Ghosal}, {Beattie}, {Ripperda}, {Porth}, \& {Bacchini}}]{Grehan2025_numerical_reconnection}
{Grehan}, M.~P., {Ghosal}, T., {Beattie}, J.~R., {et~al.} 2025, arXiv e-prints, arXiv:2503.20013, \dodoi{10.48550/arXiv.2503.20013}

\bibitem[{{Grete} {et~al.}(2021){Grete}, {O'Shea}, \& {Beckwith}}]{Grete2021_as_a_matter_of_tension}
{Grete}, P., {O'Shea}, B.~W., \& {Beckwith}, K. 2021, The Astrophysical Journal, 909, 148, \dodoi{10.3847/1538-4357/abdd22}

\bibitem[{{Grete} {et~al.}(2023){Grete}, {O'Shea}, \& {Beckwith}}]{Grete2023_as_a_matter_of_dynamical_range}
---. 2023, The Astrophysical Journal Letters, 942, L34, \dodoi{10.3847/2041-8213/acaea7}

\bibitem[{{Grete} {et~al.}(2025){Grete}, {Scannapieco}, {Br{\"u}ggen}, \& {Pan}}]{Grete2025_density_distribution}
{Grete}, P., {Scannapieco}, E., {Br{\"u}ggen}, M., \& {Pan}, L. 2025, \apj, 987, 122, \dodoi{10.3847/1538-4357/add936}

\bibitem[{{Guo} {et~al.}(2024){Guo}, {Kim}, \& {Stone}}]{Guo2024_SNe_multiphase_ISM}
{Guo}, M., {Kim}, C.-G., \& {Stone}, J.~M. 2024, arXiv e-prints, arXiv:2411.12809, \dodoi{10.48550/arXiv.2411.12809}

\bibitem[{Harris {et~al.}(2020)Harris, Millman, van~der Walt, Gommers, Virtanen, Cournapeau, Wieser, Taylor, Berg, Smith, Kern, Picus, Hoyer, van Kerkwijk, Brett, Haldane, del R{\'\i}o, Wiebe, Peterson, G{\'e}rard-Marchant, Sheppard, Reddy, Weckesser, Abbasi, Gohlke, \& Oliphant}]{numpy2020}
Harris, C.~R., Millman, K.~J., van~der Walt, S.~J., {et~al.} 2020, Nature, 585, 357, \dodoi{10.1038/s41586-020-2649-2}

\bibitem[{{Haverkorn} {et~al.}(2008){Haverkorn}, {Brown}, {Gaensler}, \& {McClure-Griffiths}}]{Haverkorn2008_turbulent_outer_scale}
{Haverkorn}, M., {Brown}, J.~C., {Gaensler}, B.~M., \& {McClure-Griffiths}, N.~M. 2008, \apj, 680, 362, \dodoi{10.1086/587165}

\bibitem[{{Hennebelle} {et~al.}(2011){Hennebelle}, {Commer{\c{c}}on}, {Joos}, {Klessen}, {Krumholz}, {Tan}, \& {Teyssier}}]{Hennebelle2011}
{Hennebelle}, P., {Commer{\c{c}}on}, B., {Joos}, M., {et~al.} 2011, Astronomy and Astrophysics, 528, A72, \dodoi{10.1051/0004-6361/201016052}

\bibitem[{{Hennebelle} \& {Iffrig}(2014)}]{Hennebelle2014_SNe_driven_clustering}
{Hennebelle}, P., \& {Iffrig}, O. 2014, \aap, 570, A81, \dodoi{10.1051/0004-6361/201423392}

\bibitem[{{Hill} {et~al.}(2012){Hill}, {Joung}, {Mac Low}, {Benjamin}, {Haffner}, {Klingenberg}, \& {Waagan}}]{Hill2012_SNe_driven_turb}
{Hill}, A.~S., {Joung}, M.~R., {Mac Low}, M.-M., {et~al.} 2012, \apj, 750, 104, \dodoi{10.1088/0004-637X/750/2/104}

\bibitem[{{Ho} {et~al.}(2024){Ho}, {Yuen}, \& {Lazarian}}]{Ho2024_multiphase_ISM_turbulence}
{Ho}, K.~W., {Yuen}, K.~H., \& {Lazarian}, A. 2024, arXiv e-prints, arXiv:2407.14199, \dodoi{10.48550/arXiv.2407.14199}

\bibitem[{{Hopkins}(2013)}]{Hopkins2013_non_lognormal_s_pdf}
{Hopkins}, P.~F. 2013, The Monthly Notices of The Royal Astronomical Society, 430, 1880, \dodoi{10.1093/mnras/stt010}

\bibitem[{{Hu}(2019)}]{Hu2019_Sne_driven_winds}
{Hu}, C.-Y. 2019, \mnras, 483, 3363, \dodoi{10.1093/mnras/sty3252}

\bibitem[{{Hu}(2025)}]{Hu2025_multiphase_ISM_turbulence}
{Hu}, Y. 2025, \apj, 986, 62, \dodoi{10.3847/1538-4357/add731}

\bibitem[{{Hu} {et~al.}(2021){Hu}, {Lazarian}, \& {Xu}}]{Hu2021_cosmic_ray_modes}
{Hu}, Y., {Lazarian}, A., \& {Xu}, S. 2021, arXiv e-prints, arXiv:2111.15066.
\newblock \doarXiv{2111.15066}

\bibitem[{Hunter(2007)}]{Hunter2007}
Hunter, J.~D. 2007, Computing in Science \& Engineering, 9, 90, \dodoi{10.1109/MCSE.2007.55}

\bibitem[{{Iroshnikov}(1964)}]{Iroshnikov_1965_IK_turb}
{Iroshnikov}, P.~S. 1964, Soviet Astronomy, 7, 566

\bibitem[{{Joblib Development Team}(2020)}]{joblib}
{Joblib Development Team}. 2020, Joblib: running Python functions as pipeline jobs.
\newblock \url{https://joblib.readthedocs.io/}

\bibitem[{{Joung} \& {Mac Low}(2006)}]{Joung2006_stratified_box}
{Joung}, M.~K.~R., \& {Mac Low}, M.-M. 2006, \apj, 653, 1266, \dodoi{10.1086/508795}

\bibitem[{{K{\"a}pyl{\"a}} {et~al.}(2018){K{\"a}pyl{\"a}}, {Gent}, {V{\"a}is{\"a}l{\"a}}, \& {Sarson}}]{Kapyla2018_helicity_SNe_turb}
{K{\"a}pyl{\"a}}, M.~J., {Gent}, F.~A., {V{\"a}is{\"a}l{\"a}}, M.~S., \& {Sarson}, G.~R. 2018, \aap, 611, A15, \dodoi{10.1051/0004-6361/201731228}

\bibitem[{{Karpov} {et~al.}(2020){Karpov}, {Martizzi}, {Macias}, {Ramirez-Ruiz}, {Kolborg}, \& {Naiman}}]{Karpov2020}
{Karpov}, P.~I., {Martizzi}, D., {Macias}, P., {et~al.} 2020, \apj, 896, 66, \dodoi{10.3847/1538-4357/ab8f23}

\bibitem[{{Kempski} {et~al.}(2023){Kempski}, {Fielding}, {Quataert}, {Galishnikova}, {Kunz}, {Philippov}, \& {Ripperda}}]{Kempski2023_b_field_reversals}
{Kempski}, P., {Fielding}, D.~B., {Quataert}, E., {et~al.} 2023, arXiv e-prints, arXiv:2304.12335, \dodoi{10.48550/arXiv.2304.12335}

\bibitem[{{Kempski} \& {Quataert}(2022)}]{Kempski2022_cr_scattering}
{Kempski}, P., \& {Quataert}, E. 2022, The Monthly Notices of The Royal Astronomical Society, 514, 657, \dodoi{10.1093/mnras/stac1240}

\bibitem[{{Kennicutt}(1998)}]{Kennicutt1998_SFR}
{Kennicutt}, Robert~C., J. 1998, \apj, 498, 541, \dodoi{10.1086/305588}

\bibitem[{{Kennicutt} {et~al.}(2007){Kennicutt}, {Calzetti}, {Walter}, {Helou}, {Hollenbach}, {Armus}, {Bendo}, {Dale}, {Draine}, {Engelbracht}, {Gordon}, {Prescott}, {Regan}, {Thornley}, {Bot}, {Brinks}, {de Blok}, {de Mello}, {Meyer}, {Moustakas}, {Murphy}, {Sheth}, \& {Smith}}]{Kennicutt2007_obsSFR}
{Kennicutt}, Robert~C., J., {Calzetti}, D., {Walter}, F., {et~al.} 2007, \apj, 671, 333, \dodoi{10.1086/522300}

\bibitem[{{Kennicutt} \& {Evans}(2012)}]{Kennicutt_MW_SB}
{Kennicutt}, R.~C., \& {Evans}, N.~J. 2012, \araa, 50, 531, \dodoi{10.1146/annurev-astro-081811-125610}

\bibitem[{{Kim} {et~al.}(2023){Kim}, {Kim}, {Gong}, \& {Ostriker}}]{Kim2023_multiphase_ISM}
{Kim}, C.-G., {Kim}, J.-G., {Gong}, M., \& {Ostriker}, E.~C. 2023, \apj, 946, 3, \dodoi{10.3847/1538-4357/acbd3a}

\bibitem[{{Kim} {et~al.}(2006){Kim}, {Kim}, \& {Ostriker}}]{Kim2006_galactic_spiral_shocks}
{Kim}, C.-G., {Kim}, W.-T., \& {Ostriker}, E.~C. 2006, \apjl, 649, L13, \dodoi{10.1086/508160}

\bibitem[{{Kim} \& {Ostriker}(2002)}]{Kim2002_MRI_galactic_disk}
{Kim}, W.-T., \& {Ostriker}, E.~C. 2002, \apj, 570, 132, \dodoi{10.1086/339352}

\bibitem[{Kochurin \& Kuznetsov(2024)}]{Kochurin_2024_acoustic}
Kochurin, E.~A., \& Kuznetsov, E.~A. 2024, Phys. Rev. Lett., 133, 207201, \dodoi{10.1103/PhysRevLett.133.207201}

\bibitem[{{Kolborg} {et~al.}(2022){Kolborg}, {Martizzi}, {Ramirez-Ruiz}, {Pfister}, {Sakari}, {Wechsler}, \& {Soares-Furtado}}]{Kolborg2022_metal_mixing_1}
{Kolborg}, A.~N., {Martizzi}, D., {Ramirez-Ruiz}, E., {et~al.} 2022, \apjl, 936, L26, \dodoi{10.3847/2041-8213/ac8c98}

\bibitem[{{Kolborg} {et~al.}(2023){Kolborg}, {Ramirez-Ruiz}, {Martizzi}, {Macias}, \& {Soares-Furtado}}]{Kolborg2023_metal_mixing_2}
{Kolborg}, A.~N., {Ramirez-Ruiz}, E., {Martizzi}, D., {Macias}, P., \& {Soares-Furtado}, M. 2023, \apj, 949, 100, \dodoi{10.3847/1538-4357/acca80}

\bibitem[{Kolmogorov(1941)}]{Kolmogorov1941}
Kolmogorov, A.~N. 1941, Doklady Akademii Nauk Sssr, 30, 301, \dodoi{10.1098/rspa.1991.0075}

\bibitem[{{Korpi} {et~al.}(1999){Korpi}, {Brandenburg}, {Shukurov}, {Tuominen}, \& {Nordlund}}]{Korpi1999_SNe_ISM}
{Korpi}, M.~J., {Brandenburg}, A., {Shukurov}, A., {Tuominen}, I., \& {Nordlund}, {\r{A}}. 1999, \apjl, 514, L99, \dodoi{10.1086/311954}

\bibitem[{{Kriel} {et~al.}(2023){Kriel}, {Beattie}, {Federrath}, {Krumholz}, \& {Hew}}]{Kriel2023_fundamental_scales_II}
{Kriel}, N., {Beattie}, J.~R., {Federrath}, C., {Krumholz}, M.~R., \& {Hew}, J. K.~J. 2023, arXiv e-prints, arXiv:2310.17036, \dodoi{10.48550/arXiv.2310.17036}

\bibitem[{{Kriel} {et~al.}(2022){Kriel}, {Beattie}, {Seta}, \& {Federrath}}]{Kriel2022_kinematic_dynamo_scales}
{Kriel}, N., {Beattie}, J.~R., {Seta}, A., \& {Federrath}, C. 2022, The Monthly Notices of The Royal Astronomical Society, 513, 2457, \dodoi{10.1093/mnras/stac969}

\bibitem[{{Krumholz} \& {Burkhart}(2016)}]{Krumholz2016_source_of_turb}
{Krumholz}, M.~R., \& {Burkhart}, B. 2016, \mnras, 458, 1671, \dodoi{10.1093/mnras/stw434}

\bibitem[{Krumholz \& McKee(2005)}]{Krumholz2005}
Krumholz, M.~R., \& McKee, C.~F. 2005, The Astrophysical Journal, 630, 250, \dodoi{10.1086/431734}

\bibitem[{{Krumholz} \& {Ting}(2018)}]{Krumholz2018_metallicity_SF}
{Krumholz}, M.~R., \& {Ting}, Y.-S. 2018, The Monthly Notices of The Royal Astronomical Society, 475, 2236, \dodoi{10.1093/mnras/stx3286}

\bibitem[{{Krumholz} {et~al.}(2025){Krumholz}, {Ting}, {Li}, {Zhang}, {Mead}, \& {Ness}}]{Krumholz2025_metal_mixing_2}
{Krumholz}, M.~R., {Ting}, Y.-S., {Li}, Z., {et~al.} 2025, arXiv e-prints, arXiv:2507.14572, \dodoi{10.48550/arXiv.2507.14572}

\bibitem[{{Kuijken} \& {Gilmore}(1989)}]{KuijkenGilmore}
{Kuijken}, K., \& {Gilmore}, G. 1989, \mnras, 239, 605, \dodoi{10.1093/mnras/239.2.605}

\bibitem[{{Kulsrud} {et~al.}(1997){Kulsrud}, {Cen}, {Ostriker}, \& {Ryu}}]{Kulsrud1997_cosmic_batteries}
{Kulsrud}, R.~M., {Cen}, R., {Ostriker}, J.~P., \& {Ryu}, D. 1997, \apj, 480, 481, \dodoi{10.1086/303987}

\bibitem[{{Lada} \& {Lada}(2003)}]{LADA_stars_in_dense}
{Lada}, C.~J., \& {Lada}, E.~A. 2003, \araa, 41, 57, \dodoi{10.1146/annurev.astro.41.011802.094844}

\bibitem[{{Lam} {et~al.}(2015){Lam}, {Pitrou}, \& {Seibert}}]{numba:2015}
{Lam}, S.~K., {Pitrou}, A., \& {Seibert}, S. 2015, in Proc. Second Workshop on the LLVM Compiler Infrastructure in HPC, 1--6, \dodoi{10.1145/2833157.2833162}

\bibitem[{{Langer} {et~al.}(2021){Langer}, {Pineda}, {Goldsmith}, {Chambers}, {Riquelme}, {Anderson}, {Luisi}, {Justen}, \& {Buchbender}}]{Langer_DIG}
{Langer}, W.~D., {Pineda}, J.~L., {Goldsmith}, P.~F., {et~al.} 2021, \aap, 651, A59, \dodoi{10.1051/0004-6361/202040223}

\bibitem[{{Liu} {et~al.}(2022){Liu}, {Qiu}, \& {Zhang}}]{Liu2021_cor_scales}
{Liu}, J., {Qiu}, K., \& {Zhang}, Q. 2022, The Astrophysical Journal, 925, 30, \dodoi{10.3847/1538-4357/ac3911}

\bibitem[{Mac~Low \& Klessen(2004)}]{MacLow2004}
Mac~Low, M.~M., \& Klessen, R.~S. 2004, Reviews of Modern Physics, 76, 125, \dodoi{10.1103/RevModPhys.76.125}

\bibitem[{{Macias} \& {Ramirez-Ruiz}(2018)}]{Macias2018}
{Macias}, P., \& {Ramirez-Ruiz}, E. 2018, \apj, 860, 89, \dodoi{10.3847/1538-4357/aac3e0}

\bibitem[{{Martizzi} {et~al.}(2015){Martizzi}, {Faucher-Gigu{\`e}re}, \& {Quataert}}]{Martizzi2015}
{Martizzi}, D., {Faucher-Gigu{\`e}re}, C.-A., \& {Quataert}, E. 2015, \mnras, 450, 504, \dodoi{10.1093/mnras/stv562}

\bibitem[{{Martizzi} {et~al.}(2016){Martizzi}, {Fielding}, {Faucher-Gigu{\`e}re}, \& {Quataert}}]{Martizzi2016}
{Martizzi}, D., {Fielding}, D., {Faucher-Gigu{\`e}re}, C.-A., \& {Quataert}, E. 2016, \mnras, 459, 2311, \dodoi{10.1093/mnras/stw745}

\bibitem[{{Matthaeus} {et~al.}(2008){Matthaeus}, {Pouquet}, {Mininni}, {Dmitruk}, \& {Breech}}]{Matthaeus2008_rapid_alignment}
{Matthaeus}, W.~H., {Pouquet}, A., {Mininni}, P.~D., {Dmitruk}, P., \& {Breech}, B. 2008, Physical Review Letters, 100, 085003, \dodoi{10.1103/PhysRevLett.100.085003}

\bibitem[{McKee \& Ostriker(2007)}]{McKee2007}
McKee, C.~F., \& Ostriker, E.~C. 2007, Annu. Rev. Astron. Astrophys., 45, 565, \dodoi{10.1146/annurev.astro.45.051806.110602}

\bibitem[{{McKee} \& {Ostriker}(1977)}]{McKee1977_ISM}
{McKee}, C.~F., \& {Ostriker}, J.~P. 1977, \apj, 218, 148, \dodoi{10.1086/155667}

\bibitem[{{McKee} {et~al.}(2015){McKee}, {Parravano}, \& {Hollenbach}}]{McKee_solarneigh}
{McKee}, C.~F., {Parravano}, A., \& {Hollenbach}, D.~J. 2015, \apj, 814, 13, \dodoi{10.1088/0004-637X/814/1/13}

\bibitem[{{Mee} \& {Brandenburg}(2006)}]{Mee2006_blastwave_turbulence}
{Mee}, A.~J., \& {Brandenburg}, A. 2006, \mnras, 370, 415, \dodoi{10.1111/j.1365-2966.2006.10476.x}

\bibitem[{{Menon} {et~al.}(2020){Menon}, {Federrath}, \& {Kuiper}}]{Menon2020b}
{Menon}, S.~H., {Federrath}, C., \& {Kuiper}, R. 2020, The Monthly Notices of The Royal Astronomical Society, 493, 4643, \dodoi{10.1093/mnras/staa580}

\bibitem[{Mocz \& Burkhart(2019)}]{Mocz2019}
Mocz, P., \& Burkhart, B. 2019, The Astrophysical Journal Letters, 884, L35, \dodoi{10.3847/2041-8213/ab48f6}

\bibitem[{Oliphant(2006)}]{Oliphant2006}
Oliphant, T. 2006, {NumPy}: A guide to {NumPy}, USA: Trelgol Publishing.
\newblock \url{http://www.numpy.org/}

\bibitem[{{Padoan} \& {Nordlund}(2002)}]{Padoan2002}
{Padoan}, P., \& {Nordlund}, {\AA}. 2002, The Astrophysical Journal, 576, 870, \dodoi{10.1086/341790}

\bibitem[{{Padoan} {et~al.}(2016){Padoan}, {Pan}, {Haugb{\o}lle}, \& {Nordlund}}]{Padoan2016_supernova_driving}
{Padoan}, P., {Pan}, L., {Haugb{\o}lle}, T., \& {Nordlund}, {\r{A}}. 2016, The Astrophysical Journal, 822, 11, \dodoi{10.3847/0004-637X/822/1/11}

\bibitem[{{Pan} {et~al.}(2016){Pan}, {Padoan}, {Haugb{\o}lle}, \& {Nordlund}}]{Luibun2016_SNe_driving_modes}
{Pan}, L., {Padoan}, P., {Haugb{\o}lle}, T., \& {Nordlund}, {\r{A}}. 2016, \apj, 825, 30, \dodoi{10.3847/0004-637X/825/1/30}

\bibitem[{pandas~development team(2023)}]{pandas}
pandas~development team, T. 2023, pandas-dev/pandas: Pandas, 2.1.4,  Zenodo, \dodoi{10.5281/zenodo.10304236}

\bibitem[{{Plunian} {et~al.}(2020){Plunian}, {Teimurazov}, {Stepanov}, \& {Verma}}]{Plunian2020_inverse_cascade}
{Plunian}, F., {Teimurazov}, A., {Stepanov}, R., \& {Verma}, M.~K. 2020, Journal of Fluid Mechanics, 895, A13, \dodoi{10.1017/jfm.2020.307}

\bibitem[{{Rosen} {et~al.}(2014){Rosen}, {Lopez}, {Krumholz}, \& {Ramirez-Ruiz}}]{Rosen2014}
{Rosen}, A.~L., {Lopez}, L.~A., {Krumholz}, M.~R., \& {Ramirez-Ruiz}, E. 2014, \mnras, 442, 2701, \dodoi{10.1093/mnras/stu1037}

\bibitem[{{Ruszkowski} \& {Pfrommer}(2023)}]{Ruszkowski2023_CRs_in_galaxies_review}
{Ruszkowski}, M., \& {Pfrommer}, C. 2023, arXiv e-prints, arXiv:2306.03141, \dodoi{10.48550/arXiv.2306.03141}

\bibitem[{{Sampson} {et~al.}(2023){Sampson}, {Beattie}, {Krumholz}, {Crocker}, {Federrath}, \& {Seta}}]{Sampson2023_SCR_diffusion}
{Sampson}, M.~L., {Beattie}, J.~R., {Krumholz}, M.~R., {et~al.} 2023, The Monthly Notices of The Royal Astronomical Society, 519, 1503, \dodoi{10.1093/mnras/stac3207}

\bibitem[{{Sampson} {et~al.}(2025){Sampson}, {Beattie}, {Teyssier}, {Kempski}, {Moseley}, {Commer{\c{c}}on}, {Dubois}, \& {Rosdahl}}]{Sampson2025_CR_transport_CRMHD}
{Sampson}, M.~L., {Beattie}, J.~R., {Teyssier}, R., {et~al.} 2025, arXiv e-prints, arXiv:2506.03768, \dodoi{10.48550/arXiv.2506.03768}

\bibitem[{{Sharda} {et~al.}(2018){Sharda}, {Federrath}, {da Cunha}, {Swinbank}, \& {Dye}}]{2018MNRAS.477.4380S}
{Sharda}, P., {Federrath}, C., {da Cunha}, E., {Swinbank}, A.~M., \& {Dye}, S. 2018, \mnras, 477, 4380, \dodoi{10.1093/mnras/sty886}

\bibitem[{Sharda {et~al.}(2021)Sharda, Krumholz, Wisnioski, Forbes, Federrath, \& Acharyya}]{Sharda2021_galactic_metallicity_modelling}
Sharda, P., Krumholz, M.~R., Wisnioski, E., {et~al.} 2021, Monthly Notices of the Royal Astronomical Society, 502, 5935, \dodoi{10.1093/mnras/stab252}

\bibitem[{{Sharda} {et~al.}(2021){Sharda}, {Menon}, {Federrath}, {Krumholz}, {Beattie}, {Jameson}, {Tokuda}, {Burkhart}, {Crocker}, {Law}, {Seta}, {Gaetz}, {Pingel}, {Seitenzahl}, {Sano}, \& {Fukui}}]{Sharda2021_driving_mode}
{Sharda}, P., {Menon}, S.~H., {Federrath}, C., {et~al.} 2021, The Monthly Notices of The Royal Astronomical Society, \dodoi{10.1093/mnras/stab3048}

\bibitem[{{Shivakumar} \& {Federrath}(2025)}]{Shivakumar2025_numerical_dissipation}
{Shivakumar}, L.~M., \& {Federrath}, C. 2025, \mnras, 537, 2961, \dodoi{10.1093/mnras/staf160}

\bibitem[{{Socrates} {et~al.}(2008){Socrates}, {Davis}, \& {Ramirez-Ruiz}}]{Socrates2008}
{Socrates}, A., {Davis}, S.~W., \& {Ramirez-Ruiz}, E. 2008, \apj, 687, 202, \dodoi{10.1086/590046}

\bibitem[{{Stone}(1991)}]{Stone_1991_runawayOB}
{Stone}, R.~C. 1991, \aj, 102, 333, \dodoi{10.1086/115880}

\bibitem[{Stroustrup(2013)}]{Stroustrup2013}
Stroustrup, B. 2013, The C++ Programming Language, 4th edn. (Addison-Wesley Professional)

\bibitem[{{Surgent} {et~al.}(2023){Surgent}, {Lopez-Rodriguez}, \& {Clark}}]{Surgent2023_magnetic_fields_in_galaxies}
{Surgent}, W.~J., {Lopez-Rodriguez}, E., \& {Clark}, S.~E. 2023, \apj, 954, 53, \dodoi{10.3847/1538-4357/ace4c0}

\bibitem[{{Sutherland} \& {Dopita}(1993)}]{SutherlandDopita}
{Sutherland}, R.~S., \& {Dopita}, M.~A. 1993, \apjs, 88, 253, \dodoi{10.1086/191823}

\bibitem[{{Teyssier}(2002)}]{Teyssier2002_ramses}
{Teyssier}, R. 2002, \aap, 385, 337, \dodoi{10.1051/0004-6361:20011817}

\bibitem[{{Theuns} {et~al.}(1998){Theuns}, {Leonard}, {Efstathiou}, {Pearce}, \& {Thomas}}]{Theuns1998_Cooling}
{Theuns}, T., {Leonard}, A., {Efstathiou}, G., {Pearce}, F.~R., \& {Thomas}, P.~A. 1998, \mnras, 301, 478, \dodoi{10.1046/j.1365-8711.1998.02040.x}

\bibitem[{{Turk} {et~al.}(2011){Turk}, {Smith}, {Oishi}, {Skory}, {Skillman}, {Abel}, \& {Norman}}]{yt}
{Turk}, M.~J., {Smith}, B.~D., {Oishi}, J.~S., {et~al.} 2011, \apjs, 192, 9, \dodoi{10.1088/0067-0049/192/1/9}

\bibitem[{{van der Velden}(2020)}]{Velden2020_cmasher}
{van der Velden}, E. 2020, The Journal of Open Source Software, 5, 2004, \dodoi{10.21105/joss.02004}

\bibitem[{van~der Walt {et~al.}(2014)van~der Walt, {S}ch\"onberger, {Nunez-Iglesias}, {B}oulogne, {W}arner, {Y}ager, {G}ouillart, {Y}u, \& the scikit-image contributors}]{vanderWalts2014}
van~der Walt, S., {S}ch\"onberger, J.~L., {Nunez-Iglesias}, J., {et~al.} 2014, PeerJ, 2, e453, \dodoi{10.7717/peerj.453}

\bibitem[{Vijayakumar {et~al.}(2025)Vijayakumar, Sun, Ostriker, Di~Teodoro, Haubner, Kim, Leroy, \& Querejeta}]{Vijayakumar_2025}
Vijayakumar, V., Sun, J., Ostriker, E.~C., {et~al.} 2025, The Astrophysical Journal, 989, 66, \dodoi{10.3847/1538-4357/ade800}

\bibitem[{{Virtanen} {et~al.}(2020){Virtanen}, {Gommers}, {Oliphant}, {Haberland}, {Reddy}, {Cournapeau}, {Burovski}, {Peterson}, {Weckesser}, {Bright}, {van der Walt}, {Brett}, {Wilson}, {Jarrod Millman}, {Mayorov}, {Nelson}, {Jones}, {Kern}, {Larson}, {Carey}, {Polat}, {Feng}, {Moore}, {Vand erPlas}, {Laxalde}, {Perktold}, {Cimrman}, {Henriksen}, {Quintero}, {Harris}, {Archibald}, {Ribeiro}, {Pedregosa}, {van Mulbregt}, \& {Contributors}}]{Virtanen2020}
{Virtanen}, P., {Gommers}, R., {Oliphant}, T.~E., {et~al.} 2020, Nature Methods, 17, 261, \dodoi{https://doi.org/10.1038/s41592-019-0686-2}

\bibitem[{Wang {et~al.}(2017)Wang, Gotoh, \& Watanabe}]{Wang_2017_3_2}
Wang, J., Gotoh, T., \& Watanabe, T. 2017, Phys. Rev. Fluids, 2, 013403, \dodoi{10.1103/PhysRevFluids.2.013403}

\bibitem[{{Westmoquette} {et~al.}(2009){Westmoquette}, {Smith}, {Gallagher}, {Trancho}, {Bastian}, \& {Konstantopoulos}}]{Westmoquette2009_velocity_dispersion_M82}
{Westmoquette}, M.~S., {Smith}, L.~J., {Gallagher}, III, J.~S., {et~al.} 2009, \apj, 696, 192, \dodoi{10.1088/0004-637X/696/1/192}

\bibitem[{{Xu} \& {Lazarian}(2022)}]{Xu2022_cosmic_ray_streaming}
{Xu}, S., \& {Lazarian}, A. 2022, The Astrophysical Journal, 927, 94, \dodoi{10.3847/1538-4357/ac4dfd}

\bibitem[{{Zakharov} \& {Sagdeev}(1970)}]{Zakharov_acoustic}
{Zakharov}, V.~E., \& {Sagdeev}, R.~Z. 1970, Soviet Physics Doklady, 15, 439

\bibitem[{{Zhang}(2018)}]{Dong2018_wind_review}
{Zhang}, D. 2018, Galaxies, 6, 114, \dodoi{10.3390/galaxies6040114}

\end{thebibliography}
\bibliographystyle{aasjournal}
\end{document}